\pdfoutput=1

\documentclass[epj,nopacs,final]{svjour}

\usepackage{graphicx}
\usepackage[switch]{lineno}
\usepackage{hyperref}
\usepackage{cite}
\usepackage{subfig}
\usepackage{multirow}
\usepackage{amsmath}
\usepackage{textcomp}

\newcommand*\patchAmsMathEnvironmentForLineno[1]{%
  \expandafter\let\csname old#1\expandafter\endcsname\csname #1\endcsname
  \expandafter\let\csname oldend#1\expandafter\endcsname\csname end#1\endcsname
  \renewenvironment{#1}%
     {\linenomath\csname old#1\endcsname}%
     {\csname oldend#1\endcsname\endlinenomath}}%
\newcommand*\patchBothAmsMathEnvironmentsForLineno[1]{%
  \patchAmsMathEnvironmentForLineno{#1}%
  \patchAmsMathEnvironmentForLineno{#1*}}%
\AtBeginDocument{%
\patchBothAmsMathEnvironmentsForLineno{equation}%
\patchBothAmsMathEnvironmentsForLineno{align}%
\patchBothAmsMathEnvironmentsForLineno{flalign}%
\patchBothAmsMathEnvironmentsForLineno{alignat}%
\patchBothAmsMathEnvironmentsForLineno{gather}%
\patchBothAmsMathEnvironmentsForLineno{multline}%
}

\journalname{Eur. Phys. J. C}

\begin{document}


\title{Final results on $^\textbf{82}$Se double beta decay to the ground state of $^\textbf{82}$Kr from the NEMO-3 experiment}


\author{
  R.~Arnold\inst{1}
  \and C.~Augier\inst{2}
  \and A.S.~Barabash\inst{3}
  \and A.~Basharina-Freshville\inst{4}
  \and S.~Blondel\inst{2}
  \and S.~Blot\inst{5}
  \and M.~Bongrand\inst{2}
  \and D.~Boursette\inst{2}
  \and V.~Brudanin\inst{6,7}
  \and J.~Busto\inst{8}
  \and A.J.~Caffrey\inst{9}
  \and S.~Calvez\inst{2}
  \and M.~Cascella\inst{4}
  \and C.~Cerna\inst{10}
  \and J.P.~Cesar\inst{11}
  \and A.~Chapon\inst{12}
  \and E.~Chauveau\inst{10}
  \and A.~Chopra\inst{4}
  \and L.~Dawson\inst{4} 
  \and D.~Duchesneau\inst{13}
  \and D.~Durand\inst{12}
  \and V.~Egorov\inst{6}
  \and G.~Eurin\inst{2,4}
  \and J.J.~Evans\inst{5}
  \and L.~Fajt\inst{14}
  \and D.~Filosofov\inst{6}
  \and R.~Flack\inst{4}
  \and X.~Garrido\inst{2}
  \and H.~G\'omez\inst{2}
  \and B.~Guillon\inst{12}
  \and P.~Guzowski\inst{5}
  \and R.~Hod\'{a}k\inst{14}
  \and A.~Huber\inst{10}
  \and P.~Hubert\inst{10}
  \and C.~Hugon\inst{10}
  \and S.~Jullian\inst{2}
  \and A.~Klimenko\inst{6}
  \and O.~Kochetov\inst{6}
  \and S.I.~Konovalov\inst{3}
  \and V.~Kovalenko\inst{6}
  \and D.~Lalanne\inst{2}
  \and K.~Lang\inst{11}
  \and Y.~Lemi\`ere\inst{12}
  \and T.~Le~Noblet\inst{13}
  \and Z.~Liptak\inst{11}
  \and X.R.~Liu\inst{4}
  \and P.~Loaiza\inst{2}
  \and G.~Lutter\inst{10}
  \and C.~Macolino\inst{2}
  \and F.~Mamedov\inst{14}
  \and C.~Marquet\inst{10}
  \and F.~Mauger\inst{12}
  \and B.~Morgan\inst{15}
  \and J.~Mott\inst{4,27}
  \and I.~Nemchenok\inst{6}
  \and M.~Nomachi\inst{16}
  \and F.~Nova\inst{11}
  \and F.~Nowacki\inst{1}
  \and H.~Ohsumi\inst{17}
  \and R.B.~Pahlka\inst{11}
  \and C.~Patrick\inst{4} 
  \and F.~Perrot\inst{10}
  \and F.~Piquemal\inst{10,18}
  \and P.~Povinec\inst{19}
  \and P.~P\v{r}idal\inst{14}
  \and Y.A.~Ramachers\inst{15}
  \and A.~Remoto\inst{13}
  \and J.L.~Reyss\inst{20}
  \and C.L.~Riddle\inst{9}
  \and E.~Rukhadze\inst{14}
  \and N.I.~Rukhadze\inst{6}
  \and R.~Saakyan\inst{4}
  \and R.~Salazar\inst{11}
  \and X.~Sarazin\inst{2}
  \and Yu.~Shitov\inst{6,21}
  \and L.~Simard\inst{2,22}
  \and F.~\v{S}imkovic\inst{19}
  \and A.~Smetana\inst{14}
  \and K.~Smolek\inst{14}
  \and A.~Smolnikov\inst{6}
  \and S.~S\"oldner-Rembold\inst{5}
  \and B.~Soul\'e\inst{10}
  \and I.~\v{S}tekl\inst{14}
  \and J.~Suhonen\inst{23}
  \and C.S.~Sutton\inst{24}
  \and G.~Szklarz\inst{2}
  \and J.~Thomas\inst{4}
  \and V.~Timkin\inst{6}
  \and S.~Torre\inst{4}
  \and Vl.I.~Tretyak\inst{25}
  \and V.I.~Tretyak\inst{6}
  \and V.I.~Umatov\inst{3}
  \and I.~Vanushin\inst{3}
  \and C.~Vilela\inst{4}
  \and V.~Vorobel\inst{26}
  \and D.~Waters\inst{4}
  \and F.~Xie\inst{4} 
  \and A.~\v{Z}ukauskas\inst{26}
}

\institute{
  IPHC, ULP, CNRS/IN2P3, F-67037 Strasbourg, France
  \and LAL, Universit\'e Paris-Sud, CNRS/IN2P3, Universit\'e Paris-Saclay, F-91405 Orsay, France
  \and NRC ``Kurchatov Institute'', ITEP, 117218 Moscow, Russia
  \and UCL, London, WC1E 6BT, United Kingdom
  \and University of Manchester, Manchester, M13 9PL,~United Kingdom
  \and JINR, 141980 Dubna, Russia
  \and National Research Nuclear University MEPhI, 115409 Moscow, Russia
  \and Aix Marseille Universit\'e, CNRS, CPPM, F-13288 Marseille, France
  \and Idaho National Laboratory, Idaho Falls, ID 83415, U.S.A.
  \and CENBG, Universit\'e de Bordeaux, CNRS/IN2P3, F-33175 Gradignan, France
  \and University of Texas at Austin, Austin, TX 78712, U.S.A.
  \and LPC Caen, ENSICAEN, Universit\'e de Caen, CNRS/IN2P3, F-14050 Caen, France
  \and LAPP, Universit\'e de Savoie, CNRS/IN2P3, F-74941 Annecy-le-Vieux, France
  \and Institute of Experimental and Applied Physics, Czech Technical University in Prague, CZ-12800 Prague, Czech Republic
  \and University of Warwick, Coventry, CV4 7AL, United Kingdom
  \and Osaka University, 1-1 Machikaneyama Toyonaka, Osaka 560-0043, Japan
  \and Saga University, Saga 840-8502, Japan
  \and Laboratoire Souterrain de Modane, F-73500 Modane, France
  \and FMFI, Comenius University, SK-842 48 Bratislava, Slovakia
  \and LSCE, CNRS, F-91190 Gif-sur-Yvette, France
  \and Imperial College London, London, SW7 2AZ, United Kingdom
  \and Institut Universitaire de France, F-75005 Paris, France
  \and Jyv\"askyl\"a University, FIN-40351 Jyv\"askyl\"a, Finland
  \and MHC, South Hadley, MA 01075, U.S.A.
  \and Institute for Nuclear Research, 03028 Kyiv, Ukraine
  \and Charles University in Prague, Faculty of Mathematics and Physics, CZ-12116 Prague, Czech Republic
  \and \emph{Present Address:} Boston University, Boston, MA 02215, U.S.A.
}

\mail{jmott@hep.ucl.ac.uk, r.saakyan@ucl.ac.uk}

\date{Date: August 26, 2018}

\abstract{
  Using data from the NEMO-3 experiment, we have measured the two-neutrino double beta decay ($2\nu\beta\beta$)
  half-life of $^{82}$Se as $T_{\smash{1/2}}^{2\nu} = \left[ 9.39 \pm 0.17\,\left(\mbox{stat}\right) \pm 0.58\,\left(\mbox{syst}\right)\right] \times 10^{19}$\,y 
  under the single-state dominance hypothesis for this nuclear transition. 
  The corresponding nuclear 
  matrix element is $\left|M^{2\nu}\right| = 0.0498 \pm 0.0016$. In addition, a search for
  neutrinoless double beta decay ($0\nu\beta\beta$) using 0.93\,kg of $^{82}$Se observed for a total of 5.25\,y
  has been conducted and no evidence
  for a signal has been found. The resulting half-life limit of $T_{1/2}^{0\nu} > 2.5 \times 10^{23} \,\mbox{y} \,(90\%\,\mbox{C.L.})$
  for the light neutrino exchange mechanism leads to a constraint on the effective Majorana neutrino mass of
  $\langle m_{\nu} \rangle < \left(1.2 - 3.0\right) \,\mbox{eV}$, where the range reflects $0\nu\beta\beta$ nuclear matrix
  element values from different calculations. Furthermore, constraints on lepton number violating parameters
  for other $0\nu\beta\beta$ mechanisms, such as right-handed currents, majoron emission and R-parity violating
  supersymmetry modes have been set.
}

\maketitle

\section{Introduction}
The observation of neutrino oscillations has provided proof that the neutrino has non-zero mass \cite{Fukuda:1998mi,Ahmad:2002jz,Agashe:2014kda}.
However the absolute mass of the neutrino and its fundamental Dirac or Majorana nature remain undetermined.
Neutrinoless double beta decay ($0\nu\beta\beta$) is the only practical way to establish the full lepton number violation
required by many grand unification models and if the decay proceeds via a light neutrino exchange mechanism, would be
one of the most sensitive probes of absolute neutrino mass \cite{Dell'Oro:2016dbc}. 

The half-life of $0\nu\beta\beta$ is given by: 
\begin{equation}
[T_{1/2}^{0\nu}]^{-1} = G^{0\nu} g^{4}_{A} \vert M^{0\nu} \vert ^{2} \langle \xi \rangle^{2} \,,
\label{equ:0vbb}
\end{equation}
where $g_A $ is the axial-vector coupling constant,
$G^{0\nu}$ is a phase-space factor, $M^{0\nu}$ is a nuclear matrix element (NME)
and $\langle \xi \rangle$ is a lepton number violating parameter. In the most commonly discussed mechanism
of $0\nu\beta\beta$, the decay proceeds via the exchange of a light Majorana neutrino 
($\langle \xi \rangle \equiv \langle m_{\nu} \rangle/ m_e$, where $m_e$ is the mass of the electron).
However, other mechanisms are possible, such as the admixture of right-handed currents in the electroweak
interaction, majoron emission and R-parity violating supersymmetry (SUSY).
In all mechanisms, $0\nu\beta\beta$ violates lepton number conservation and is a direct probe of physics beyond the
Standard Model. To date, no evidence for $0\nu\beta\beta$ has been found, with the best half-life limits
in the $10^{24}-10^{26}$\,y range \cite{Gando:2016pfg,Agostini:2018tnm,Alduino:2017ehq,Arnold:2015wpy,Azzolini:2018dyb,Albert:2017owj,Aalseth:2017btx}. 

Two-neutrino double beta decay ($2\nu\beta\beta$) is a rare second order process that is allowed in the
Standard Model. It has been observed in 12 isotopes with half-lives ranging from 10$^{19}$ to 10$^{24}$\,y
\cite{Saakyan:2013rvw,Barabash:2015eza}. Measurement of the $2\nu\beta\beta$ half-life provides experimental
determination of the NME for this process, $M^{2\nu}$, which can be used to improve NME calculations for the
$0\nu\beta\beta$ mode. The precision with which $\langle \xi \rangle$ can be measured depends crucially on
knowledge of $M^{0\nu}$. In addition, $2\nu\beta\beta$ is an irreducible background component to $0\nu\beta\beta$
and therefore precise measurements of $2\nu\beta\beta$ rates and spectral shapes are important. 

One of the most promising double beta decay ($\beta\beta$) candidates is $^{82}$Se due to its high $Q$-value
(2997.9(3)\,keV \cite{Lincoln:2012fq}), above most common backgrounds from natural radioactivity, relatively high
isotopic abundance (8.83\% \cite{Meija:2016xx}) and existing robust technologies of isotopic enrichment through
centrifugation. It has been selected as the isotope of choice for a number of planned $0\nu\beta\beta$ decay
experiments \cite{Arnold:2010tu,Beeman:2013ahp}. 

The first measurement of $\beta\beta$ in $^{82}$Se was made in 1967 with a geochemical experiment,
extracting a half-life of $\left( 0.6^{+0.6}_{-0.3} \right) \times 10^{20}$\,y \cite{Kirsten:1967}. This result was later
confirmed by many other geochemical measurements (see reviews \cite{Kirsten:1986uh,Manuel:1986ug,Manuel:1991zz}).
Such geochemical experiments are not able to distinguish between $0\nu\beta\beta$ and $2\nu\beta\beta$ modes and
the conclusion that $2\nu\beta\beta$ had been observed was drawn using complementary theoretical and experimental
arguments.  Whilst the precision of any individual measurement was reasonably good, the spread of the results was
quite high. Nevertheless, the combination of many experiments led to a half-life value of
$\left(1.0 - 1.3\right) \times 10^{20}$\,y \cite{Manuel:1986ug,Kirsten:1986uh,Manuel:1991zz}.

The isotope of $^{82}$Se was in fact the first nucleus
in which $2\nu\beta\beta$ was directly observed  in a counter experiment in 1987 \cite{Moe:1987prl}. 
A total of 36 candidate $2\nu\beta\beta$ events were observed yielding a half-life of 
$1.1^{+0.8}_{-0.3} \times 10^{20}$\,y. A more precise direct measurement was later carried out by NEMO-2,
$\left[8.3 \pm 1.0\,(\mbox{stat})\right.$ $\left.\pm \,0.7\,(\mbox{syst})\right] \times 10^{19}$\,y \cite{Arnold:1998npa}. 
The most precise result to date was obtained by NEMO-3 after analysing a subset of its data,
$\left[9.6 \pm 0.3\,(\mbox{stat}) \pm 1.0\,(\mbox{syst})\right] \times 10^{19}$\,y \cite{Arnold:2005rz}. The same data set 
was also used to obtain a stringent lower limit on the half-life for the $0\nu\beta\beta$ decay
of $^{82}$Se, $T_{1/2}^{0\nu} > 1.0 \times 10^{23}$\,y at 90\% C.L. 
  
We present the results of the $^{82}$Se $2\nu\beta\beta$ measurement and $0\nu\beta\beta$ searches
with the full data set collected by the NEMO-3 detector, representing a five-fold increase in exposure compared
to the previously published result \cite{Arnold:2005rz}.
 
\section{\texorpdfstring{NEMO-3 Detector and $^{82}$Se Source}{NEMO-3 Detector and 82-Se Source}}
NEMO-3 was a detector composed of a tracker and a calorimeter capable of reconstructing the full topology
of $\beta\beta$ events. It was installed in the Modane Underground Laboratory (LSM) with an overburden of
4800\,m.w.e. to shield against cosmic rays. The detector housed seven enriched $\beta\beta$ isotopes in the form of thin
(about $50\,\mbox{mg}/\mbox{cm}^2$) source foils. These were arranged in a cylindrical geometry subdivided into 20
identical sectors.  The two isotopes with the largest mass were $^{100}$Mo (6.91\,kg) and $^{82}$Se (0.93\,kg)
with smaller quantities of $^{48}$Ca, $^{96}$Zr, $^{116}$Cd, $^{130}$Te and $^{150}$Nd
\cite{Arnold:2015wpy,Arnold:2005rz,Arnold:2016ezh,Argyriades:2009ph,Arnold:2011gq,Arnold:2016dpe}. Charged particle
ionisation tracks are reconstructed from hits in 50\,cm deep and 270\,cm long wire chambers on each side of the source
foils composed of 6180 Geiger cells operating in helium with the addition of ethanol as a quencher (4\%), argon (1\%) and 
water vapour (0.15\%). The transverse and longitudinal resolution of individual tracker cells was 0.5\,mm and 8.0\,mm
($\sigma$) respectively. The tracker was enclosed by calorimeter walls assembled from plastic scintillator
blocks coupled to low background photomultipliers (PMT).  The detector was calibrated by deploying $^{207}$Bi, $^{90}$Sr and $^{232}$U
sources during the course of data collection. The energy resolution of the calorimeter blocks was $5.8-7.2\%$ and the time resolution was
250\,ps, both $\sigma$ at 1\,MeV. The detector was surrounded by a solenoid which generated a 25\,G magnetic field parallel to the cell wires.
The magnetic field allows the rejection of approximately 95\% of positrons at 1\,MeV. The detector was placed in passive shielding
consisting of a 19\,cm thick layer of iron to suppress the external gamma ray background, as well as borated
water, paraffin and wood to moderate and absorb the environmental neutron background. A detailed description of
the detector and its calibration and performance can be found in \cite{Arnold:2004xq,Arnold:2015wpy}.

The $^{82}$Se source foils had a composite structure. Enriched  $^{82}$Se powder was mixed with 
polyvinyl alcohol (PVA) glue and deposited between 23\,{\textmu}m (2.2\,mg/cm$^2$) thick Mylar foils. Enriched selenium from two
production runs was used, attaining enrichment factors of $97.02\pm0.05\%$ for run 1 and $96.82\pm0.05\%$ for run 2.
Selenium from run 1, which was also used in the NEMO-2 experiment \cite{Arnold:1998npa}, was placed in a single
detector sector, while the isotope from run 2 was in an adjacent sector. The total mass of the $^{82}$Se isotope in
NEMO-3 was $\left(0.932 \pm 0.005\right)\,\mbox{kg}$, with $0.464\,\mbox{kg}$ from run 1 and $0.468\,\mbox{kg}$ from run 2.

NEMO-3 took data from February 2003 to January 2011.  A standard set of criteria define high quality runs,
where the detector was operating stably and the calorimeter was calibrated \cite{Arnold:2015wpy}.  The accepted live-time of the
detector is 5.252\,y, resulting in an exposure of 4.90\,kg$\cdot$y for $^{82}$Se.

During the first 18 months of data-taking, the radon ($^{222}$Rn) level inside the detector was higher than anticipated. 
This was caused by the diffusion of radon from the air of the laboratory into the tracking gas.  To lower the radon level inside
the detector, an anti-radon tent containing filtered air was built around the detector reducing the radon level
in the tracker volume by a factor of about 6 \cite{Argyriades:2009vq}. The higher radon activity data-taking period, lasting
1.06\,y, is referred to as phase 1 and the lower activity period, with a duration of 4.19\,y, as phase 2.

\section{Particle Identification and Event Selection}
\label{sec:ParticleID}
One of the major strengths of the NEMO-3 approach amongst $\beta\beta$ experiments is its ability 
to use multiple observables and a combination of tracking and calorimetry information 
for particle identification and reconstruction of different event topologies.
By separating data events into different channels based on the number of electrons, $\gamma$-rays and
$\alpha$-particles that they contain, a pure $\beta\beta$ signal channel can be defined along with a series
of background channels that may be used to normalise the different background contributions to this signal channel.

Electrons and positrons are identified by ionisation traces that can be extrapolated to an energy deposit in the
calorimeter, and are distinguished by their curvature in the magnetic field. By contrast, $\gamma$-rays are
identified as an energy deposit in the calorimeter without an associated track. A 1\,MeV photon has a 50\% probability
of interaction with a scintillator block. Therefore neighbouring calorimeter hits are clustered together and
attributed to a single $\gamma$-ray interaction event with an energy equal to the energy sum of the individual hits.
Due to their heavy ionisation energy losses, $\alpha$-particles from radioactive decays can not travel more than
about 35\,cm in the NEMO-3 tracker and are identified by their short, straight tracks.
 
Both data and Monte Carlo simulations (MC) of signal and background are processed by the same reconstruction
algorithm. The DECAY0 event generator \cite{Ponkratenko:2000um} is used for generation of initial kinematics and particles are
tracked through a detailed GEANT3 based detector simulation \cite{Brun:1987ma}. 

Candidate $\beta\beta$ signal events are selected to contain two electron tracks, each with an energy deposit
$> 300\,\mbox{keV}$. The tracks must originate from the $^{82}$Se source foil and have a common vertex
(i.e. the distance between the track intersections with the foil should be $\Delta_{XY} < 2$\,cm (transversely) and 
$\Delta_{Z} < 4$\,cm (vertically), set by the resolution of the tracking detector). There should be no $\alpha$-particle
tracks in the event. The timing of the calorimeter hits must be consistent with an \emph{internal} event
defined as two electrons simultaneously emitted from a common vertex in the foil \cite{Arnold:2015wpy}.

\begin{figure*}
  \subfloat[Before Removal]{\includegraphics[width=0.5\textwidth]{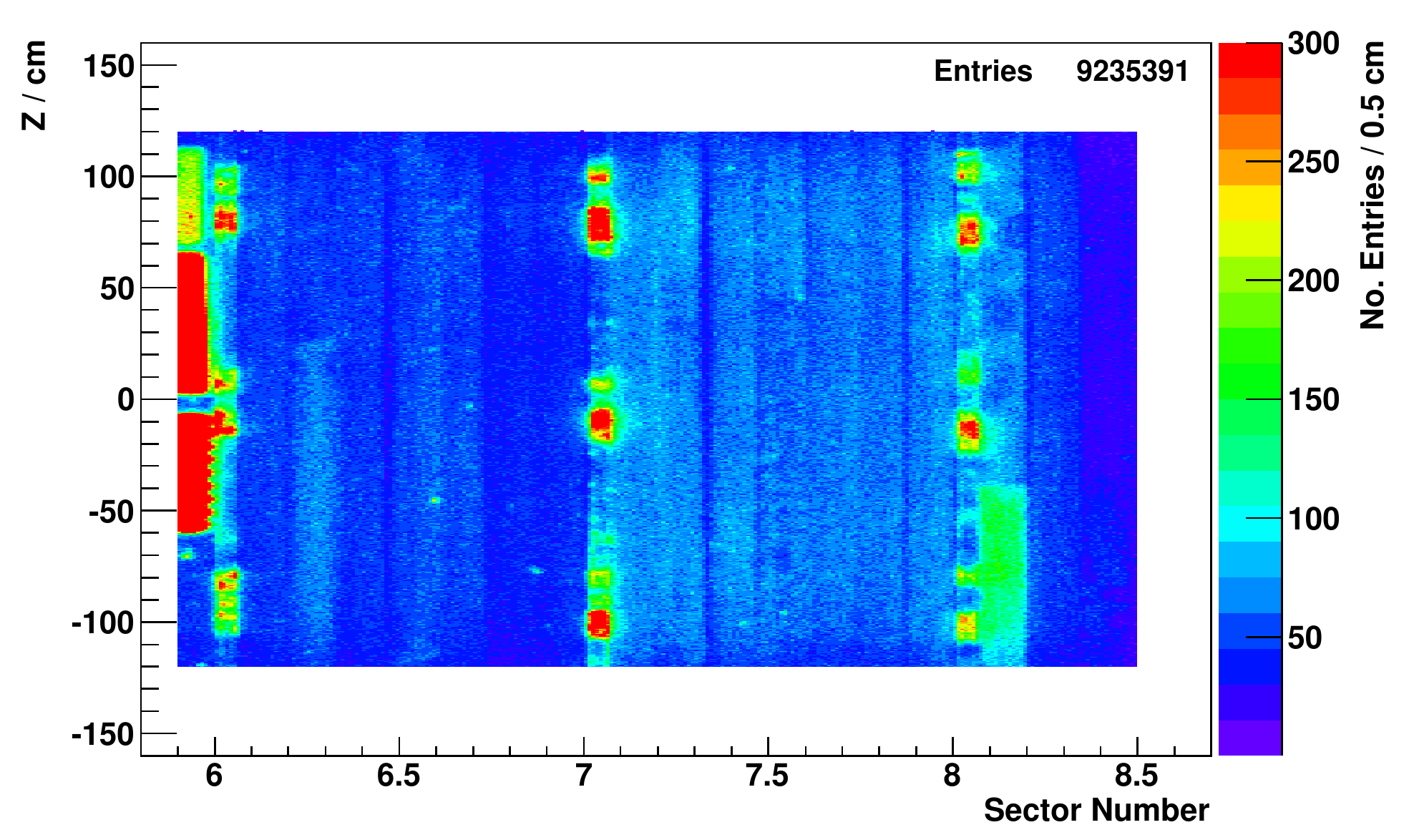} \label{fig:HotSpotsBefore}}\hfill%
  \subfloat[After Removal] {\includegraphics[width=0.5\textwidth]{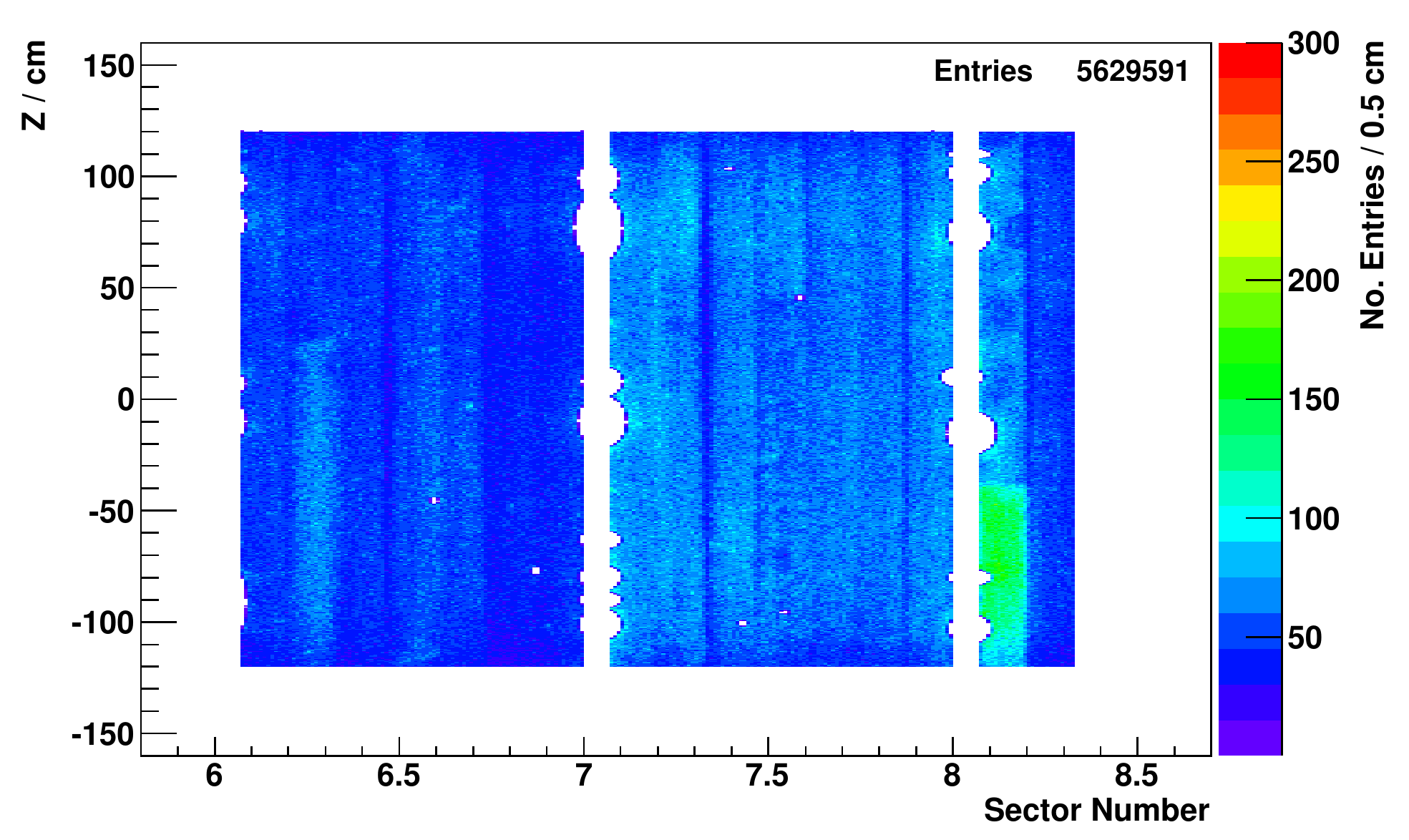} \label{fig:HotSpotsAfter}}\\
  \caption{The sectors of the detector containing the $^{82}$Se source foils, imaged in the 1e channel.  Figure~(a) shows the reconstructed vertex from all single electron events. Figure~(b) shows the same region after removing events originating from neighbouring foils, calibration tubes and areas with activity more than 5$\sigma$ higher than the mean for the particular $^{82}$Se foil strip. This is the fiducial area used in the analysis. The higher activity strip in sector 8 is contaminated with $^{210}$Bi which does not affect the $\beta\beta$ analyses due to its low $Q$-value.}
  \label{fig:HotSpots}
\end{figure*}

Backgrounds are constrained using specific event topologies and timing characteristics. Single electron
candidate events (1e) must have one electron track originating from a $^{82}$Se source foil. The position of
these intersections are used to identify areas in the source foils with higher than average contaminations
as shown in Figure~\ref{fig:HotSpots}. Areas with an event rate more than 5$\sigma$ higher than the mean rate
for the foil strip in which it is housed are excluded from the data analysis. 

The 1e1$\alpha N\gamma$ channel events contain a single electron track and a delayed $\alpha$-particle track emitted
from a common vertex, with no constraints on the number of $\gamma$-rays present. The $\alpha$-particle track must be
registered in the range of $\left(10 - 650\right)$\,{\textmu}s after the electron track, such that it is consistent
with $^{214}$Bi $\rightarrow$ $^{214}$Po $\rightarrow$ $^{210}$Pb sequential decays from the $^{238}$U radioactive series. These decays
predominantly originate from radon in the tracker as outlined in Section~\ref{sec:Backgrounds}.

Events with a single electron track and a number of $\gamma$-ray hits (1e$N\gamma$) are used to constrain
different backgrounds depending on the number of $\gamma$-rays and their timing characteristics. As with 
electron candidates, $\gamma$-ray hits must have an energy deposit $> 300$\,keV to be accepted. Events
containing electron and $\gamma$-ray hits consistent with simultaneous emission from the same location in
a $^{82}$Se foil are used to measure internal contamination by radioactive isotopes.  Conversely, those
containing hit times consistent with a $\gamma$-ray first interacting with a calorimeter block before
producing an electron in the foil are used to measure the external $\gamma$-ray flux.
Finally, crossing-electron events, where a single electron crosses from one side of the detector to the other,
are selected using the same cuts as for the $\beta\beta$ channel but with a requirement that the timing of
the calorimeter hits be consistent with an external origin of the event.
Further details on using topological, timing and energy cuts for background identification 
can be found in \cite{Argyriades:2009vq}.

\section{Background and Control Measurements}
\label{sec:Backgrounds}
Any event containing two reconstructed electrons from an origin other than the decay of $^{82}$Se can be
misidentified as a $\beta\beta$ event. The main source of background events are trace amounts of
naturally-occurring radioactive isotopes that come from the $^{238}$U and $^{232}$Th radioactive series. 
Only $\left( \beta, \gamma \right)$-emitting radioactive isotopes with high $Q$-values are potential backgrounds to a $0\nu\beta\beta$ search. 
The two main isotopes of concern are $^{214}$Bi and $^{208}$Tl with $Q$-values of $3.27$ and $4.99$\,MeV respectively. 

The largest background contribution comes from \emph{internal} contamination of the source foils. Isotopes that undergo
$\beta$-decay can mimic two electron events via the processes of $\beta$-decay with M{\o}ller scattering, $\beta$-decay
to an excited state followed by internal conversion, or by subsequent Compton scattering of the de-excitation photon.

Other background events may be classified as coming from an origin \emph{external} to the source foils. These usually involve
a $\gamma$-ray that interacts with the source foil causing pair production, Compton interaction followed by M{\o}ller scattering or double
Compton scattering.  The sources of external $\gamma$-rays are predominantly radioactive decays within the rock surrounding
the laboratory, neutron capture and decays within the detector components or shielding.

A subset of the external backgrounds is identified as \emph{radon} backgrounds, coming from $^{222}$Rn, which is a gaseous
isotope in the $^{238}$U chain. Due to its long half-life of 3.82 days 
$^{222}$Rn can be introduced via a number of mechanisms, notably emanation from detector
materials, contamination of the tracker gas or of other detector surfaces, or via diffusion through detector seals.
Once inside the detector, the radon decays to predominantly positive ions. These charged progenies drift towards
the source foils or tracker wires where they settle, leaving deposits of $^{214}$Bi near the source material \cite{Argyriades:2009vq}.
Once on or near the source foils, this $^{214}$Bi is then capable of producing background events in the same way as internal
contaminants.

The background model is defined by the activity of each isotope in specific locations.  In all background sources, $^{214}$Pb
is assumed to be in secular equilibrium with $^{214}$Bi and likewise for $^{228}$Ac, $^{212}$Bi and $^{208}$Tl.  The fitting
procedure extracts the different isotope activities using a binned log-likelihood maximisation. The distributions from
the six background channels (1e, 1e1$\alpha N \gamma$, 1e1$\gamma$, 1e2$\gamma$, external 1$\gamma$1e and crossing-electron)
 and a $\beta\beta$ signal channel are fitted simultaneously to extract the most likely activity parameters.

\subsection{External backgrounds}
The external $\gamma$-ray flux incident on the detector is quantified using the external 1$\gamma$1e and crossing-electron
channels.  In the former, a $\gamma$-ray deposits energy in the calorimeter before interacting with the source
foil to produce an outgoing electron.  In the latter, the $\gamma$-ray interacts close to the surface of a calorimeter, producing
an electron that crosses the whole tracking chamber including the source foil.  Data from these channels constrain the number of
events in the $\beta\beta$ channel from the external $\gamma$-ray flux.

The external background model is an \emph{effective model} of the $\gamma$-ray flux incident on the detector, with components 
similar to the model in \cite{Argyriades:2009vq}. It is dominated by $^{40}$K, $^{208}$Tl and $^{214}$Bi contamination in the
calorimeter PMT glass and by $^{208}$Tl, $^{214}$Bi and $^{60}$Co in the iron shielding surrounding the detector.

\begin{figure*}
  \subfloat[External 1$\gamma$1e]{\includegraphics[width=0.5\textwidth]{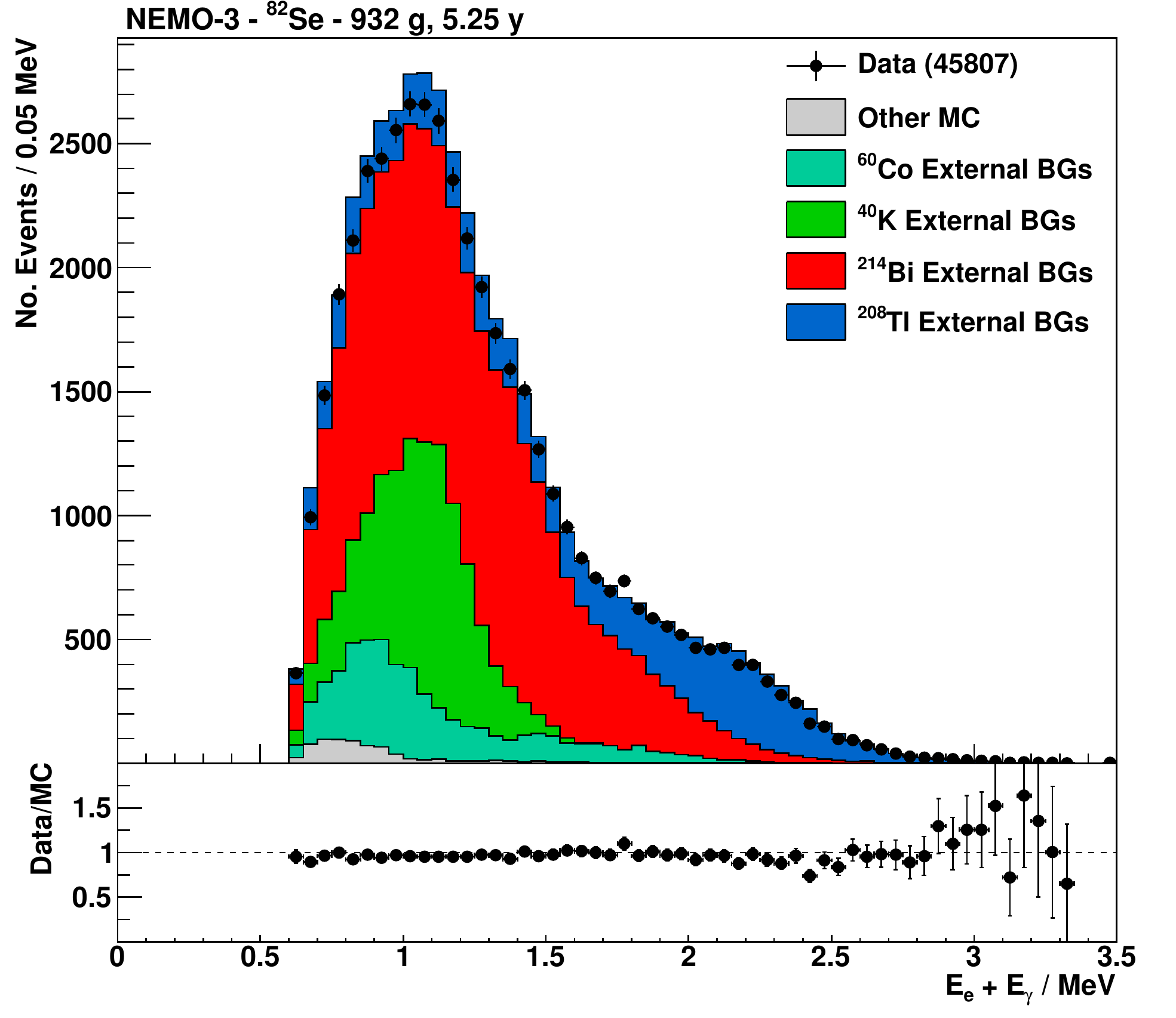}}\hfill%
  \subfloat[Crossing-electron] {\includegraphics[width=0.5\textwidth]{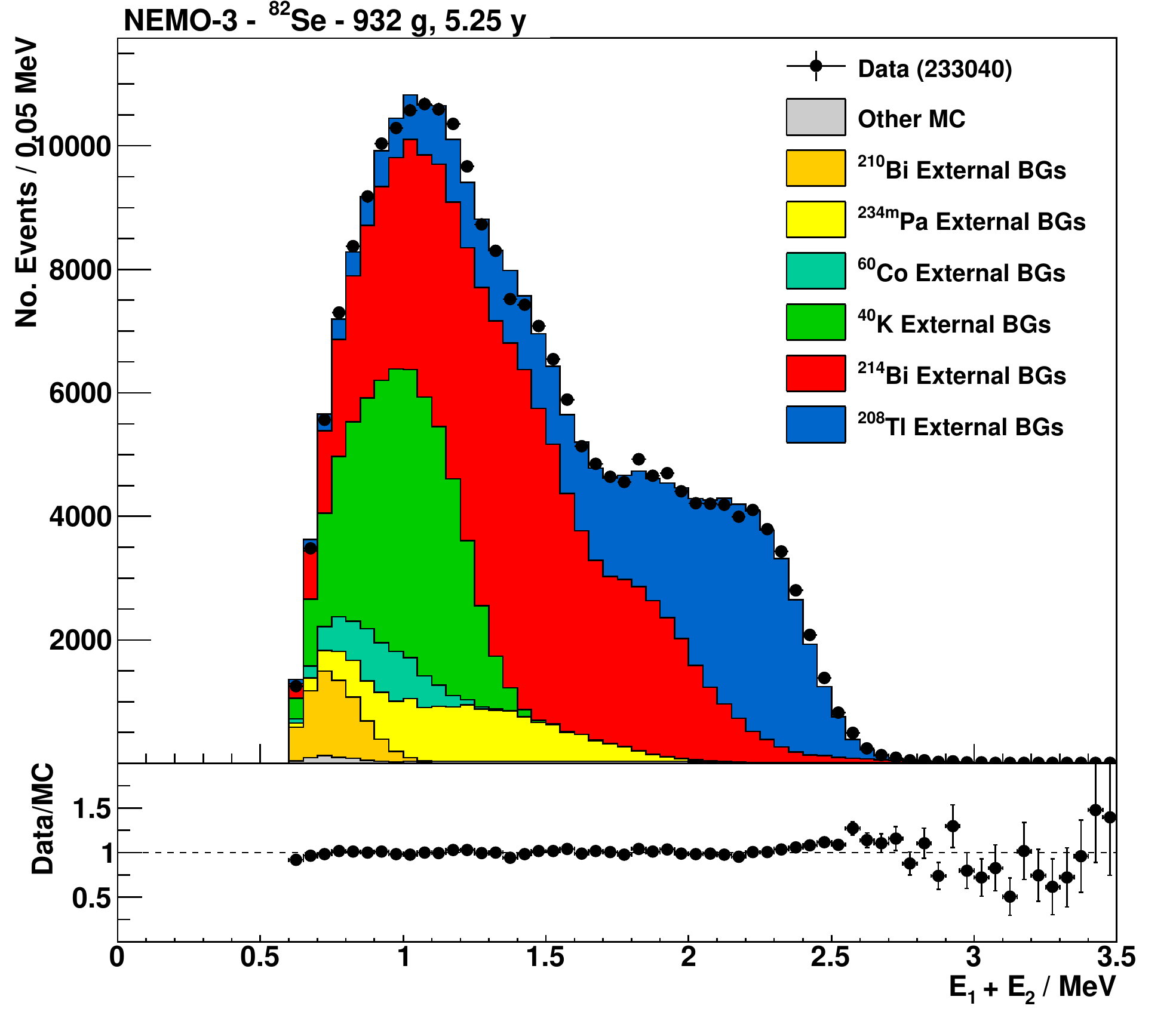}}\\
  \caption{(a) Distribution of the sum of the energies of the electron and $\gamma$-ray in the external 1$\gamma$1e channel. (b) Distribution of the sum of the incoming and outgoing electron energies in the crossing-electron channel. In both figures, the energy spectra from data are compared to the total MC prediction (top panels) and as a ratio of data to the total MC prediction (bottom panels). The Other MC histograms contain the small contributions from internal and radon background sources.}
  \label{fig:ExtBGs}
\end{figure*}

The model reproduces the data accurately as can be seen from the distributions of energy deposited in the calorimeter for the
external 1$\gamma$1e and crossing-electron channels shown in Figure~\ref{fig:ExtBGs}.

The external background model presented here is constructed using data from the $^{82}$Se sectors only. It is consistent 
with the average external background model in \cite{Argyriades:2009vq}, where all sectors are used, within $10\%-20\%$.  This is
the expected level of sector-to-sector variation in the external background model. 

\subsection{Radon backgrounds}
The radon level inside the detector can be measured by studying $^{214}$Bi $\rightarrow$ $^{214}$Po $\rightarrow$ $^{210}$Pb sequential decay events
in the 1e1$\alpha N \gamma$ channel. The distribution of the length of the $\alpha$-particle tracks is used to reconstruct the location of $^{214}$Bi. 
For example, the $\alpha$ track length is sensitive to whether the $\alpha$-particle originated from the
surface of a tracker wire or inside the bulk of the source foil.

Using the reconstructed position of the events, an extensive radon model has been developed with $^{214}$Bi on the surface of the
tracker wires, source foils and scintillators varying from sector-to-sector and, in the case of the surface of the wires, with
tracker layer \cite{Argyriades:2009vq}.

\begin{figure*}
  \subfloat[Phase 1]{\includegraphics[width=0.5\textwidth]{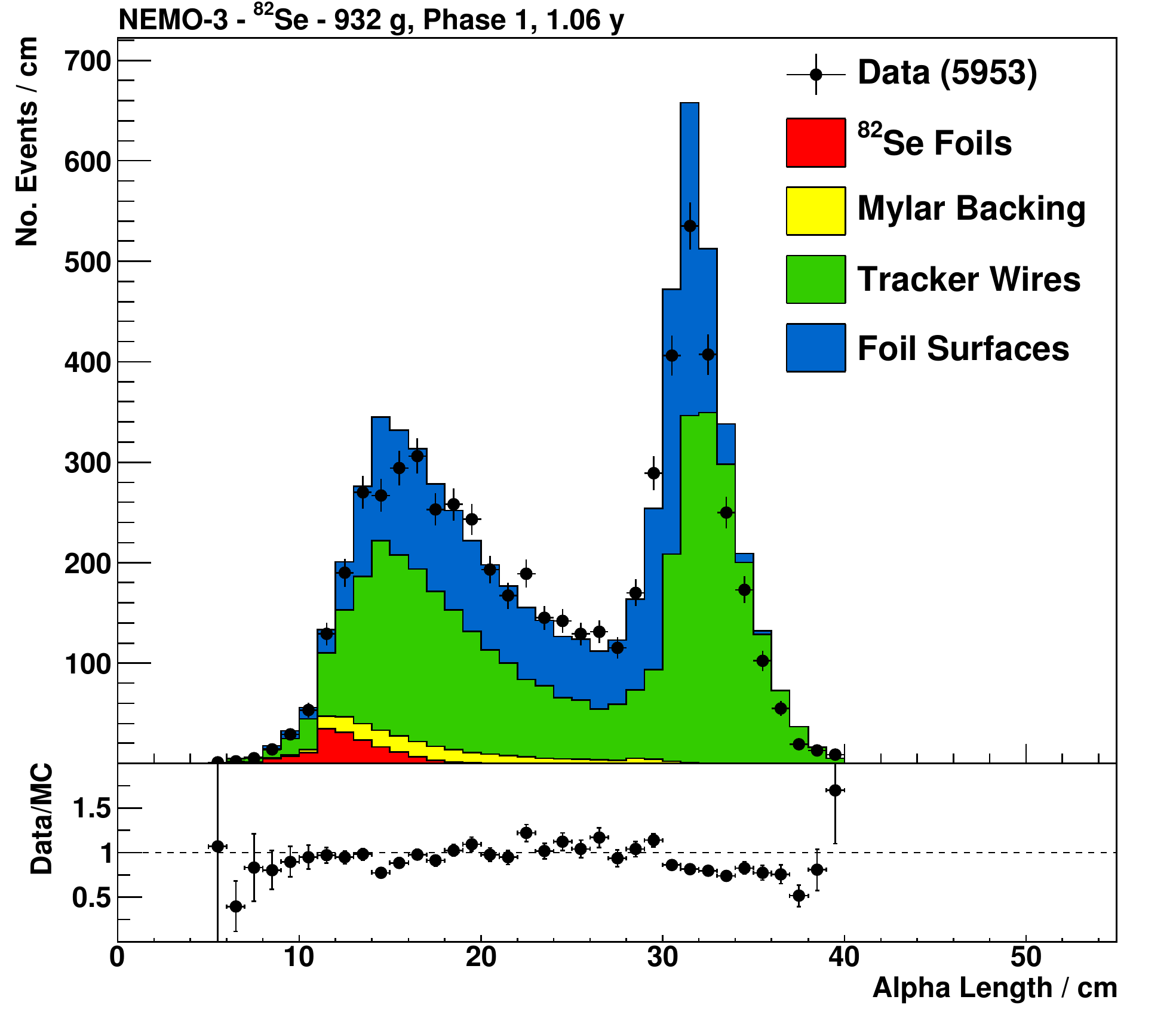}}\hfill%
  \subfloat[Phase 2]{\includegraphics[width=0.5\textwidth]{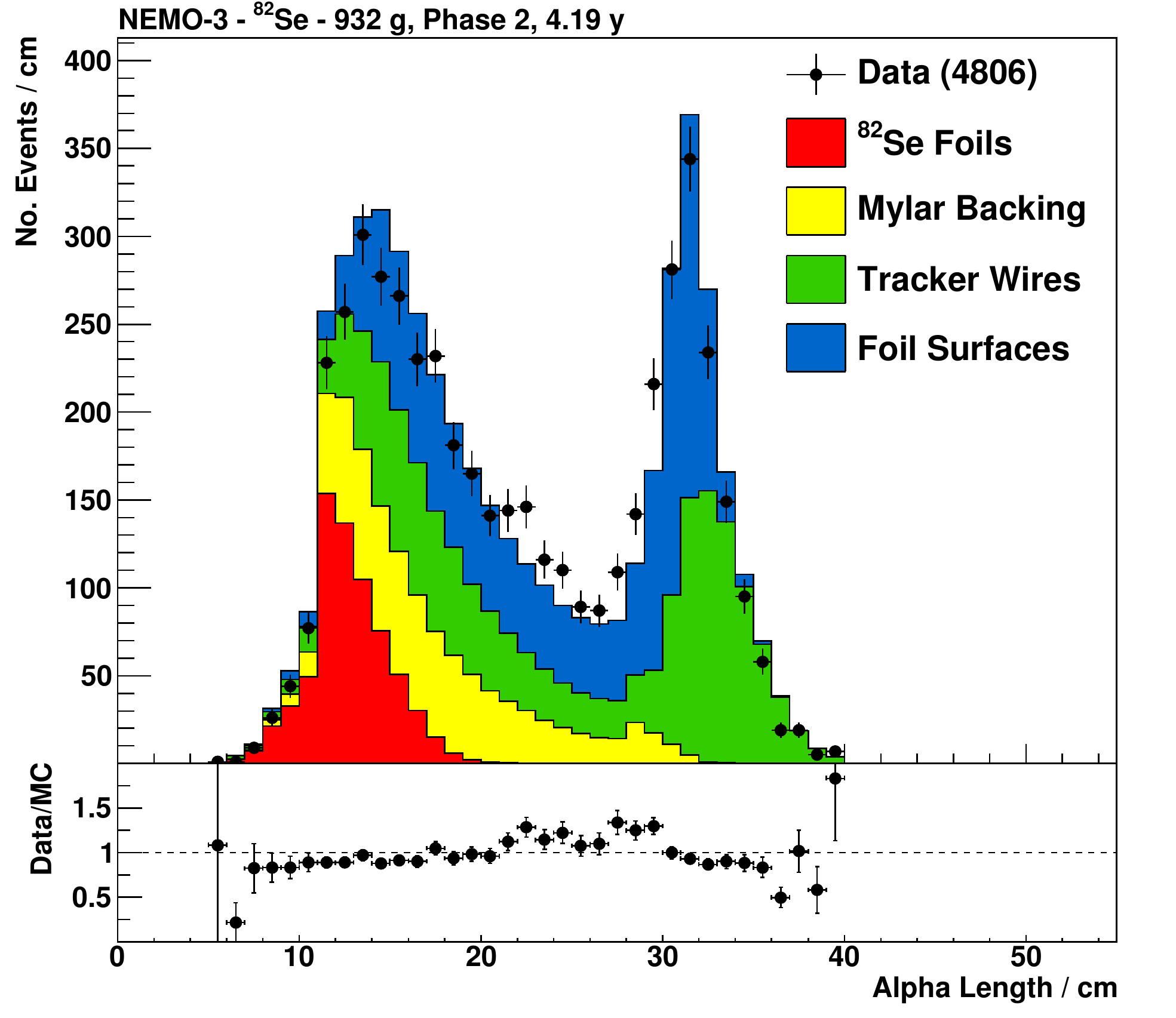}}\\
  \caption{Distributions of the length of $\alpha$-particle tracks from the 1e1$\alpha N \gamma$ channel, which contains events with one electron track and one delayed $\alpha$-particle track, with no constraints on the number of $\gamma$-rays present. The length is measured as the distance from the electron vertex on the foil to the furthest hit in the $\alpha$-particle track. Figure (a) shows data from phase 1, which had a higher radon level in the tracker and Figure (b) shows the same distribution for phase 2 data. In the top panels, data are overlaid on stacked histograms of the MC prediction from $^{214}$Bi contaminations in the source foils (red), Mylar backing film (yellow), deposits on the tracker wires (green) and on the surface of the source foils (blue). The activities of the source foil and Mylar film contaminations are the same in both phases. The bottom panels show the ratio of data to the total MC prediction.}
  \label{fig:OneEOneA}
\end{figure*}

Distributions of $\alpha$-particle track length from the 1e1$\alpha N \gamma$ channel, which are used to extract the $^{214}$Bi activities,
can be seen in Figure~\ref{fig:OneEOneA}.  The contribution from internal foil contamination has the shortest track lengths 
as these $\alpha$-particles must traverse the most material before entering the tracking gas 
while the surface of tracker wires sample has the longest tracks. 
The shape of the distributions is an artefact of the tracker geometry. The lower number of events between 20 and 30\,cm is a result of
a gap in the layers of tracker cells at this distance due to the presence of calorimeter blocks in the detector end caps \cite{Arnold:2004xq}.

The difference between phases 1 and 2 is apparent, with a higher proportion of events from surfaces of the tracker wires and source foils
during phase 1. In these cases, $^{214}$Bi has been deposited on exposed surfaces as a result of radon decay in the tracker gas.  In phase 2
there is a larger contribution from the internal and Mylar components.  This originates from $^{214}$Bi decays from contamination with
$^{226}$Ra and has therefore remained constant whilst the radon level inside the tracker gas has decreased.

The small discrepancies observed between MC and data distributions are due to a strong sensitivity of the $\alpha$-particle range to the
location of the $^{214}$Bi.  For example, the distributions can be altered significantly  by transferring $^{214}$Bi between the surface
of the foils and the surface of the wires or between different wires within the tracker.  The detection efficiency for electrons from 
$^{214}$Bi is much less sensitive to these small changes in decay location and so the systematic uncertainty from this discrepancy that
propagates through to the $\beta\beta$ channel is negligible.

In addition to the $^{214}$Bi components that are measured with $^{214}$Bi $\rightarrow$ $^{214}$Po $\rightarrow$ $^{210}$Pb delayed events, 
there are other background events from $^{208}$Tl and $^{210}$Bi. The former is a product of $^{220}$Rn decay and was measured using
1e2$\gamma$ and 1e3$\gamma$ channels where the electron track starts away from the foil \cite{Argyriades:2009vq}.  
The latter is caused by $^{210}$Pb from $^{222}$Rn deposited on the surfaces of detector components during construction.
This isotope has a half-life of $22.3\,\mbox{y}$ and supplies $^{210}$Bi over the lifetime of the experiment. It is therefore not in
equilibrium with $^{222}$Rn observed in the detector.  In a similar manner to the $^{214}$Bi activities, a map of relative $^{210}$Bi
activities divided by sector and tracker layer has been developed \cite{Argyriades:2009vq}.

\subsection{Internal backgrounds}

\begin{figure*}
  \subfloat[1e Channel]         {\includegraphics[width=0.5\textwidth]{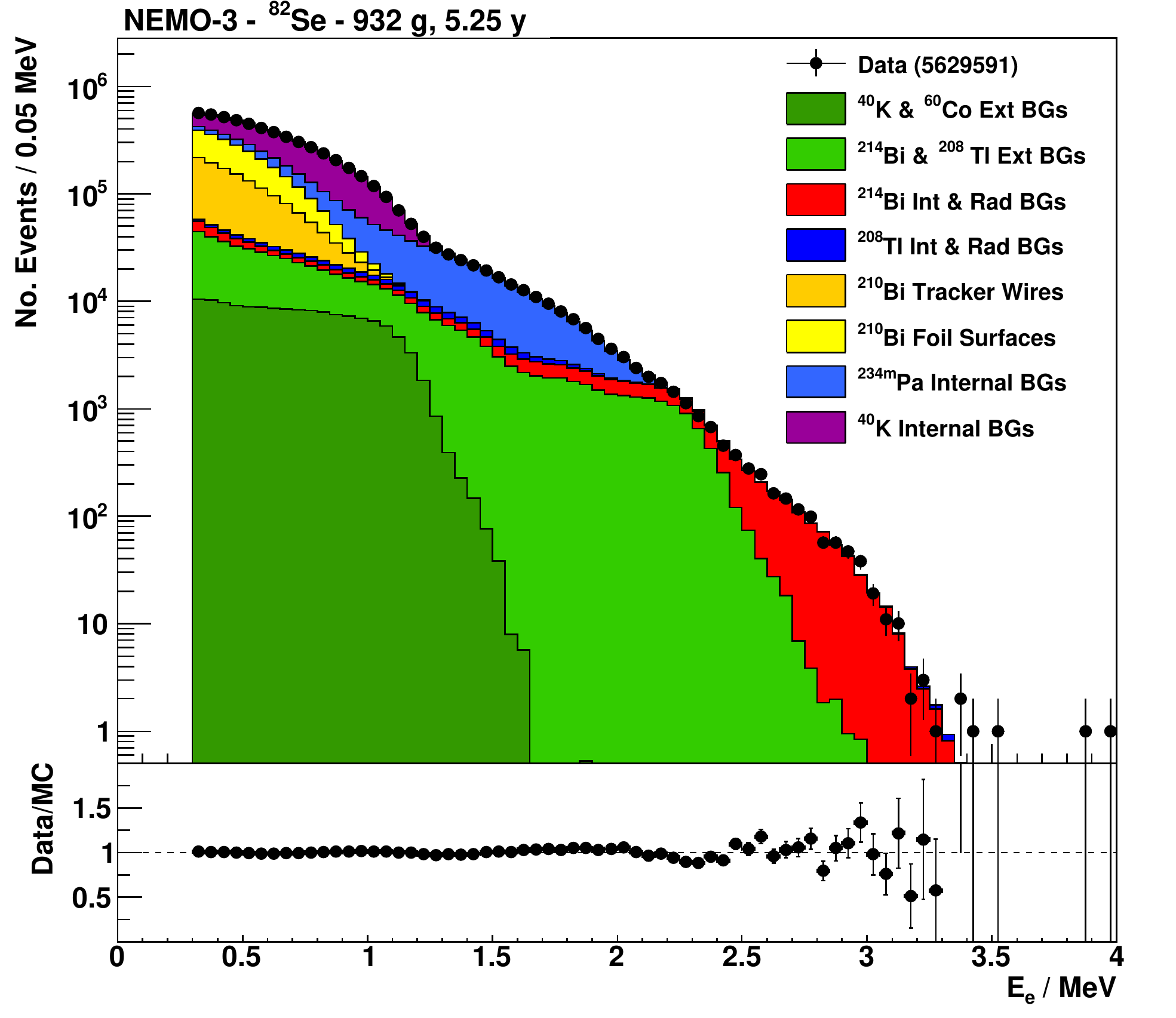}  \label{fig:OneE}}\hfill%
  \subfloat[1e1$\gamma$ Channel]{\includegraphics[width=0.5\textwidth]{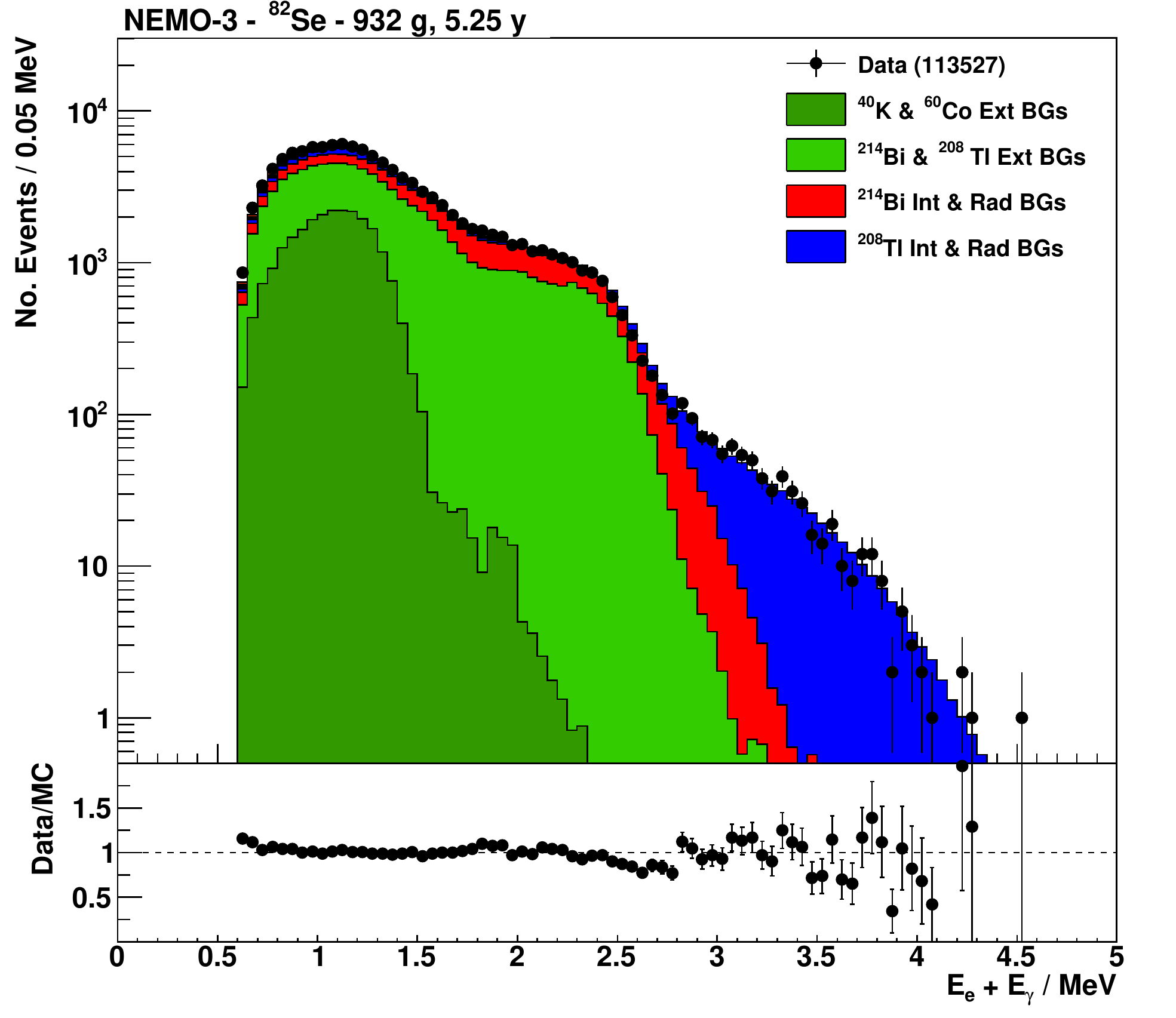}\label{fig:OneEOneG}}\\
  \caption{(a) Energy spectrum of electrons in events containing a single electron selected in the 1e channel compared to the MC prediction.  At energies below $1\,\mbox{MeV}$ the dominant contributions are from $^{40}$K contamination and from $^{210}$Bi deposited on the surfaces of the source foils and tracker wires. The contribution from  $^{234\text{m}}$Pa becomes significant in an intermediate region at $1.2-2.2\,\mbox{MeV}$. Above this energy the spectrum is composed of events from $^{214}$Bi originating from source foil contamination and radon backgrounds. The small numbers of events at the highest energies are a result of external neutron flux. They do not contribute significantly to any other channel. (b) Distribution of the sum of electron and $\gamma$-ray energies for events selected in the 1e1$\gamma$ channel comparing data to the MC prediction. Below $2.5\,\mbox{MeV}$ the distributions are mainly composed of events from the external $\gamma$-ray flux.  Above this energy the spectrum contains events from contamination of the source foil with $^{208}$Tl.}
\end{figure*}

The main backgrounds in the low energy region come from $\beta$-decaying isotopes. The 1e channel electron
energy distributions, shown in Figure~\ref{fig:OneE}, are dominated by $^{210}$Bi, $^{40}$K and $^{234\text{m}}$Pa.
In the higher energy region, the contributions from the external
$^{208}$Tl and $^{214}$Bi backgrounds become significant and at energies above $2.7\,\mbox{MeV}$, 
$^{214}$Bi from the internal and surface of tracker wire contaminations are the only remaining contributions.

The 1e1$\gamma$ channel constrains isotopes decaying to excited states, most notably $^{214}$Bi and
$^{208}$Tl as shown in Figure~\ref{fig:OneEOneG}.  At energies below $2.5\,\mbox{MeV}$ the channel serves as a
cross-check on the number of external $\gamma$-ray flux events that have calorimeter timings consistent with an
event of internal origin.  At high energies, the distribution contains events from internal contamination with $^{208}$Tl.

\begin{figure}[h]
  \includegraphics[width=0.5\textwidth]{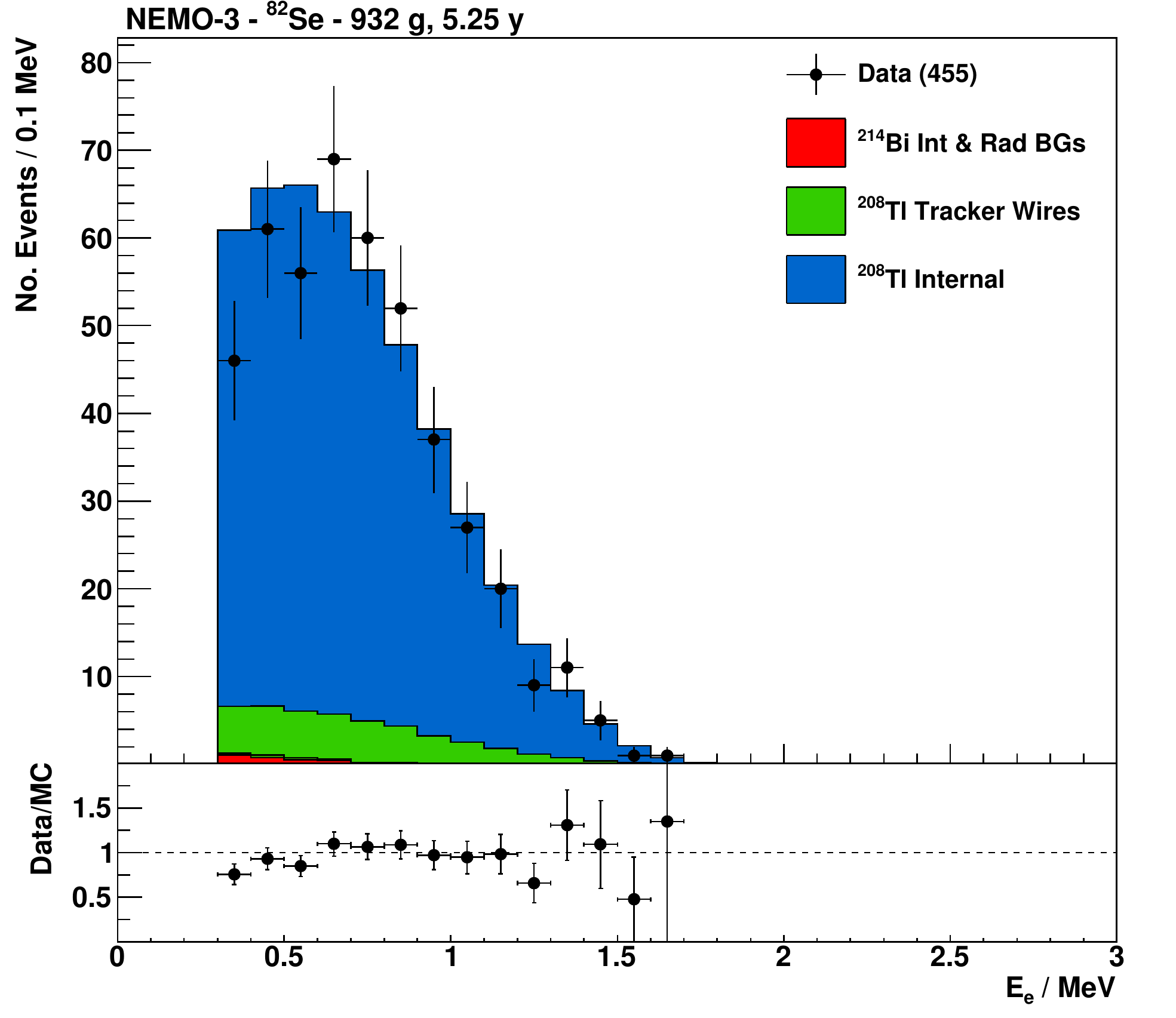}
  \caption{Energy spectrum of electrons selected in the 1e2$\gamma$ channel, which contains an electron and two $\gamma$-rays. The channel is designed to allow a measurement of the source foil contamination with $^{208}$Tl. Data are compared to the MC prediction, which is dominated by internal $^{208}$Tl contamination with a small contribution from $^{208}$Tl in the tracker wires.} 
  \label{fig:OneETwoG}
\end{figure}

A more sensitive probe for the $^{208}$Tl internal contamination is the 1e2$\gamma$ channel with one
$\gamma$-ray above 1.7\,MeV, shown in Figure~\ref{fig:OneETwoG}. Any contributions from $^{214}$Bi are
heavily suppressed by this cut on the $\gamma$ energy such that the channel is dominated by the internal
contributions of $^{208}$Tl, with a 10\% contribution from the $^{208}$Tl in the tracker wires.  

The measured activities for the internal contaminations are summarised in Table~\ref{tab:InternalHPGe}.
The levels of contamination are similar for both enrichment runs with the exception of $^{234\text{m}}$Pa where there is a
four-fold increase in the activity in run 2.
The results are compared with measurements made with a high purity germanium (HPGe) detector 
carried out prior to the installation of the $^{82}$Se foils in the detector. The results are consistent across all isotopes in Table 1. 

\begin{table}
  \caption{Measurements of the specific activity of $^{82}$Se source foils for different isotopes, made with the NEMO-3 detector and independently with an HPGe detector. Equal numbers of $^{82}$Se foils from enrichment runs 1 and 2 were measured together in the HPGe detector, so the NEMO-3 combined values are the mean values of the specific activities from each enrichment run.  All error bars are statistical only and are at the $1\sigma$ level.  The limit shown is at the $2\sigma$ level.  The HPGe measurements are taken from \cite{Arnold:2004xq}.}
  \label{tab:InternalHPGe}
  \begin{tabular*}{\columnwidth}{@{\extracolsep{\fill}}ccccc@{}}
    \hline\noalign{\smallskip}
    \multirow{2}{*}{Isotope} & \multicolumn{3}{c}{NEMO-3 (mBq/kg)}     & HPGe\\
                             & Run 1 & Run 2 & Combined & (mBq/kg)\\
    \noalign{\smallskip}\hline\noalign{\smallskip}
    $^{214}$Bi         & $1.57 \pm 0.05$ & $1.42 \pm 0.05$ & $1.50 \pm 0.04$ & $1.2 \pm 0.5$\\
    $^{208}$Tl         & $0.34 \pm 0.01$ & $0.44 \pm 0.01$ & $0.39 \pm 0.01$ & $0.4 \pm 0.1$\\
    $^{234\text{m}}$Pa & $7.5\pm 0.1$    & $27.0 \pm 0.1$  & $17.3\pm 0.1$   & $<18$\\
    $^{40}$K           & $58.1\pm 0.1$   & $59.3 \pm 0.2$  & $58.7\pm 0.1$   & $55 \pm 5$\\
    \noalign{\smallskip}\hline
  \end{tabular*}
\end{table}

\section{Two-neutrino double beta decay}
\label{sec:TwoETwoNu}
Candidate $\beta\beta$ signal events are selected using the criteria outlined in Section~\ref{sec:ParticleID}. A total of $8936$ candidate events were selected,
with $4350$ and $4586$ from source foils from enrichment runs 1 and 2 respectively.
Table~\ref{tab:TwoEEvents} shows the contribution expected from simulations of individual background sources to the $\beta\beta$ signal channel, 
with the lower energy threshold column relevant to a $2\nu\beta\beta$ measurement.

\begin{table*}
  \caption{Predicted number of events in the two electron channel for different event sources.  The expected numbers of events in the energy region relevant to 0$\nu\beta\beta$ are also given.  Isotopes denote the internal contaminations of the $^{82}$Se foils and the numbers are assigned to whether the tracks originated from foils from enrichment run 1 or 2. Multiple isotopes listed on the same line indicates the assumption of secular equilibrium.}
  \label{tab:TwoEEvents}
  \begin{tabular*}{\textwidth}{@{\extracolsep{\fill}}lcccccc@{}}
    \hline\noalign{\smallskip}
    \multirow{3}{*}{Event Source}      & \multicolumn{4}{c}{$E_{tot}>0.6\,\mbox{MeV}$}                         & \multicolumn{2}{c}{$2.6\,\mbox{MeV}<E_{tot}< 3.2\,\mbox{MeV}$}\\
                                       & \multicolumn{2}{c}{Expected Events} & \multicolumn{2}{c}{\% of Total} & Expected Events & \% of Total\\
                                       & Run 1  & Run 2                      & Run 1 & Run 2                   & Run 1 \& Run 2  & Run 1 \& Run 2\\
    \noalign{\smallskip}\hline\noalign{\smallskip}
    $^{214}$Bi, $^{214}$Pb             &  $102.2 \pm 2.7$  &   $86.2 \pm 2.2$   &   2.4 &   1.9                   & $4.1 \pm 0.1$ & 34\\
    $^{208}$Tl, $^{212}$Bi, $^{228}$Ac &   $56.8 \pm 1.3$  &   $65.1 \pm 1.4$   &   1.3 &   1.4                   & $3.1 \pm 0.1$ & 25\\
    $^{234\text{m}}$Pa                 &  $341.7 \pm 2.1$  & $1061.4 \pm 6.6$   &   7.9 &  23.0                   & $<0.1$\\
    $^{40}$K                           &  $154.1 \pm 2.2$  &  $131.6 \pm 1.8$   &   3.5 &   2.8                   & $<0.1$\\
    Radon                              &   $86.2 \pm 2.6$  &   $75.6 \pm 2.3$   &   2.0 &   1.6                   & $3.0 \pm 0.1$ & 25\\
    External                           &  $133.6 \pm 8.6$  &  $123.5 \pm 8.0$   &   3.1 &   2.7                   & $0.1 \pm 0.1$ & 1\\
    \hline\noalign{\smallskip}                                        
    All Backgrounds                    &  $874.6 \pm 7.6$  & $1543.4 \pm 13.3$  &  20.1 &  33.4                   & $10.3 \pm 0.1$ & 84\\
    $^{82}$Se 2$\nu\beta\beta$ Signal  & $3472 \pm 49$ & $3079 \pm 43$  &  79.9 &  66.6                   & $1.9 \pm 0.1$ & 16\\
    \noalign{\smallskip}\hline\noalign{\smallskip}                    
    Signal + Background                & $4347 \pm 45$ & $4622 \pm 48$  & 100.0 & 100.0                   & $12.2 \pm 0.2$ & 100\\
    Data                               &   4350            &   4586             &   N/A &   N/A                   & 15   & N/A\\
    \noalign{\smallskip}\hline
  \end{tabular*}
\end{table*}

The largest background contribution comes from internal contamination of the source foils with 15.1\% of the total number of events for run 1 foils and 29.1\% of those from run 2 foils.
Among the internal contaminants,  $^{234\text{m}}$Pa is the most prominent, accounting for 7.9\% of events originating in run 1 foils and 23.0\% of events from run 2 foils.
The external backgrounds account for 3\% of the total with the majority of events from $\gamma$-ray transitions of $^{208}$Tl and $^{214}$Bi.  
The radon backgrounds make up 2\% with a dominant contribution from $^{214}$Bi, and a secondary contribution from $^{210}$Bi.  
The majority of these events come from the surface of the tracker wires, but some are also present on the surface of the foil.  
There are more expected radon background events in phase 1 compared to phase 2 despite its much shorter exposure period.

\begin{figure*}
  \subfloat[Enrichment Run 1 Foils]{\includegraphics[width=0.5\textwidth]{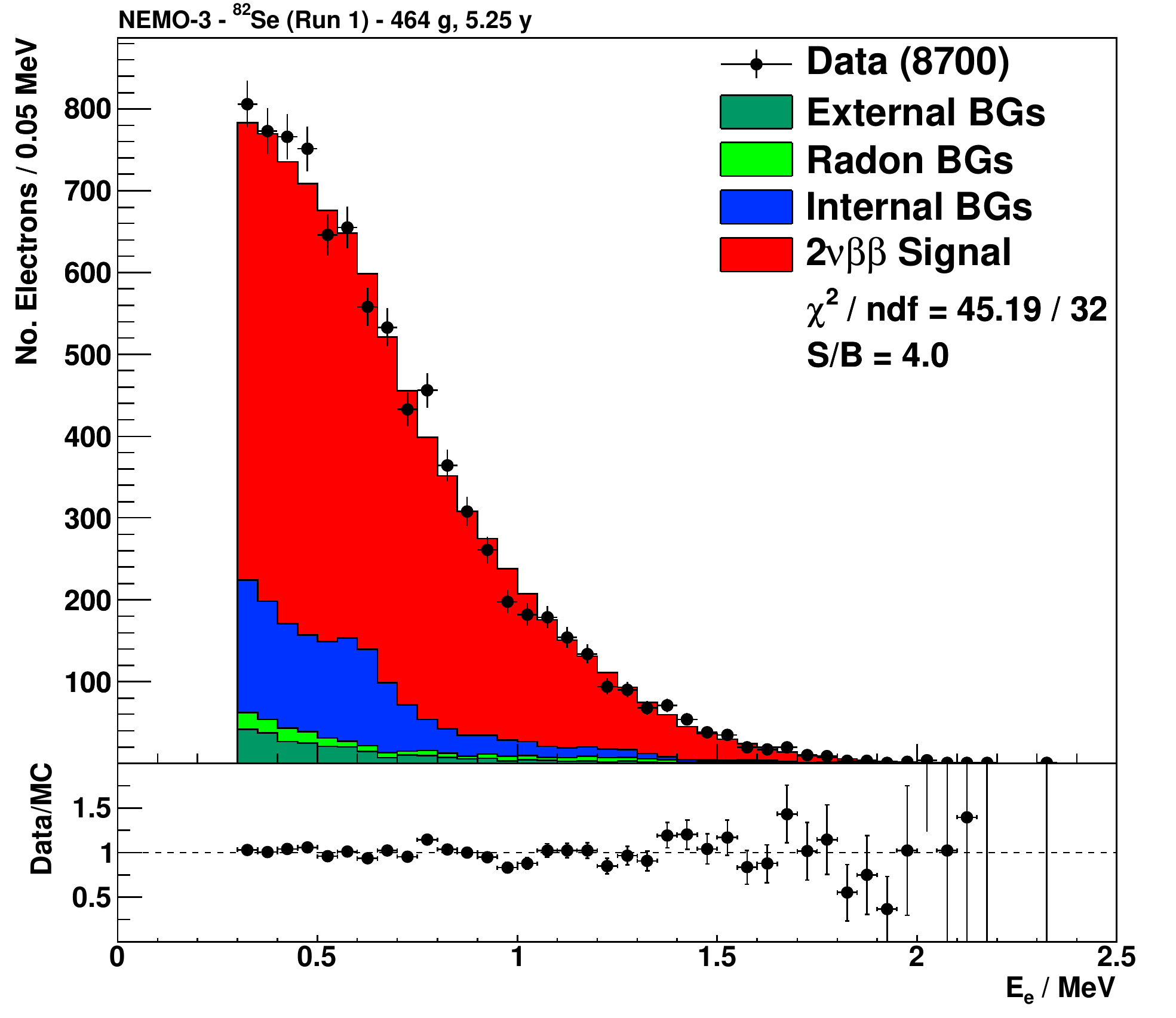} \label{fig:SingleEEnergySeOld}}\hfill%
  \subfloat[Enrichment Run 2 Foils]{\includegraphics[width=0.5\textwidth]{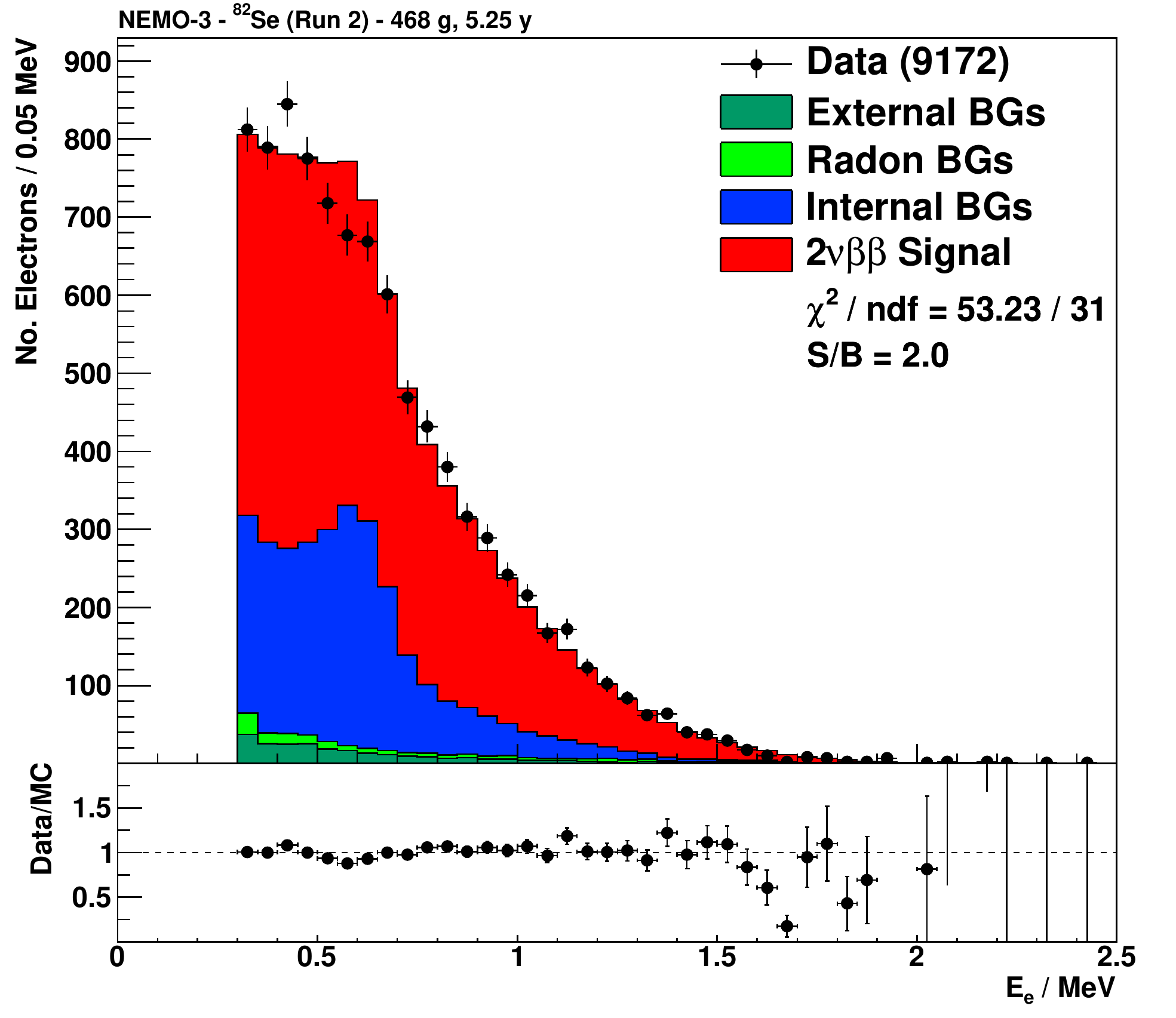} \label{fig:SingleEEnergySeNew}}\\
  \caption{Energy distribution of individual electrons in the $\beta\beta$ channel for foils from each enrichment run, showing a comparison of the data to the predicted spectrum from MC. The two electrons in each event are entered separately into this distributions. The higher level of $^{234\text{m}}$Pa contamination in the run 2 foils leads to a larger contribution from the internal backgrounds.  These foils are removed from the 2$\nu\beta\beta$ analysis due to poor modelling of this isotope (see text).}
  \label{fig:SingleEEnergy}
\end{figure*}

NEMO-3 has the unique capability of reconstructing the full kinematics of the $\beta\beta$ decay final states.  The individual energies of each electron can be seen in Figure~\ref{fig:SingleEEnergy}, where the higher degree of contamination from $^{234\text{m}}$Pa in the run 2 foils leads to a much larger contribution from the internal backgrounds.
There is a discrepancy between data and MC in the region of $0.5-0.7$\,MeV caused by a peak from 
the emission of a  694\,keV internal conversion electron from $^{234\text{m}}$Pa. This discrepancy is significantly stronger 
in the run 2 foils due to their higher contamination with  $^{234\text{m}}$Pa. The discrepancy is most likely caused by 
inaccuracies in the internal conversion electron transition probabilities obtained from the existing 
nuclear data sheets \cite{Akovali:1994xx,Browne:2007yy}. Given this large uncertainty associated with the $^{234\text{m}}$Pa
background contribution, the enrichment run 2 foils are excluded from the analysis 
to enable a more precise measurement of the $2\nu\beta\beta$ half-life, as further discussed in 
\ref{ssec:HalfLife}.

\subsection{Higher-state vs single-state dominated transistions}
For the purpose of the nuclear matrix element calculation, the decay of $^{82}$Se to $^{82}$Kr is 
modelled as two virtual $\beta$ transitions: one between the ground state of $^{82}$Se and 
the $1^+$ states of the intermediate nucleus of $^{82}$Br, and one between the $1^+$ states of $^{82}$Br and the ground state of $^{82}$Kr. 
If one single intermediate $1^+$ state dominates the transition, then the process is said to be single-state dominated (SSD). Alternatively, if the process proceeds through many higher intermediate excited states, it is said to be higher-state dominated (HSD).  
Previously, it has been assumed that $^{82}$Se decay occurs in the HSD scenario. However, a strong transition in the $^{82}$Se($^3$He,$^3$H)$^{82}$Br reaction via the 
$1^+$ (75\,keV) excited level of $^{82}$Br was recently identified \cite{Frekers:2016xx}, 
suggesting that the SSD scenario could be realised. The shape of the distribution of the 
sum of electron energies, which is used for the $2\nu\beta\beta$ half-life measurement, 
is very similar in both scenarios. However, the sub-division of energy between the electrons is different 
in the two cases and therefore a precise high-statistics study of single-electron energy distributions 
can be used to distinguish between the two models \cite{Simkovic:2005xx}. 
Moreover, the choice of the model affects the measured half-life of the $2\nu\beta\beta$ transition. 
This is because the increased number of lower energy electrons in the SSD model reduces the detection 
efficiency and therefore the extracted half-life.
The selection efficiency for the $2\nu\beta\beta$ signal calculated from MC 
using the event selection criteria described above is 
$\left[2.971 \pm 0.002\,(\mbox{stat})\right]\%$ under the HSD hypothesis and 
$\left[2.623 \pm 0.002\,(\mbox{stat})\right]\%$ in the SSD case.

\begin{figure*}
  \subfloat[Higher-state Dominated (HSD)]{\includegraphics[width=0.5\textwidth]{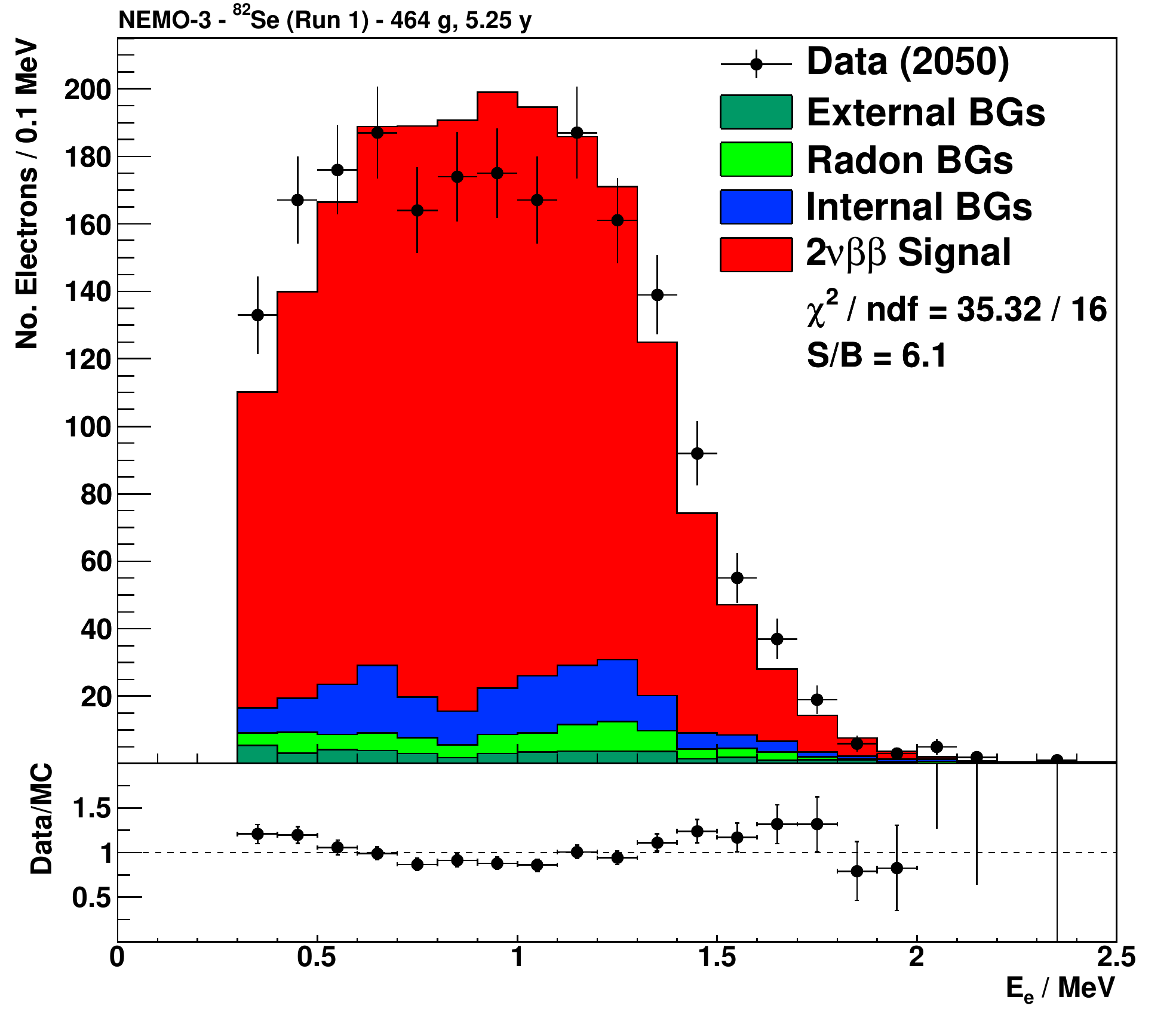} \label{fig:ESumCutHSD}}\hfill%
  \subfloat[Single-state Dominated (SSD)]{\includegraphics[width=0.5\textwidth]{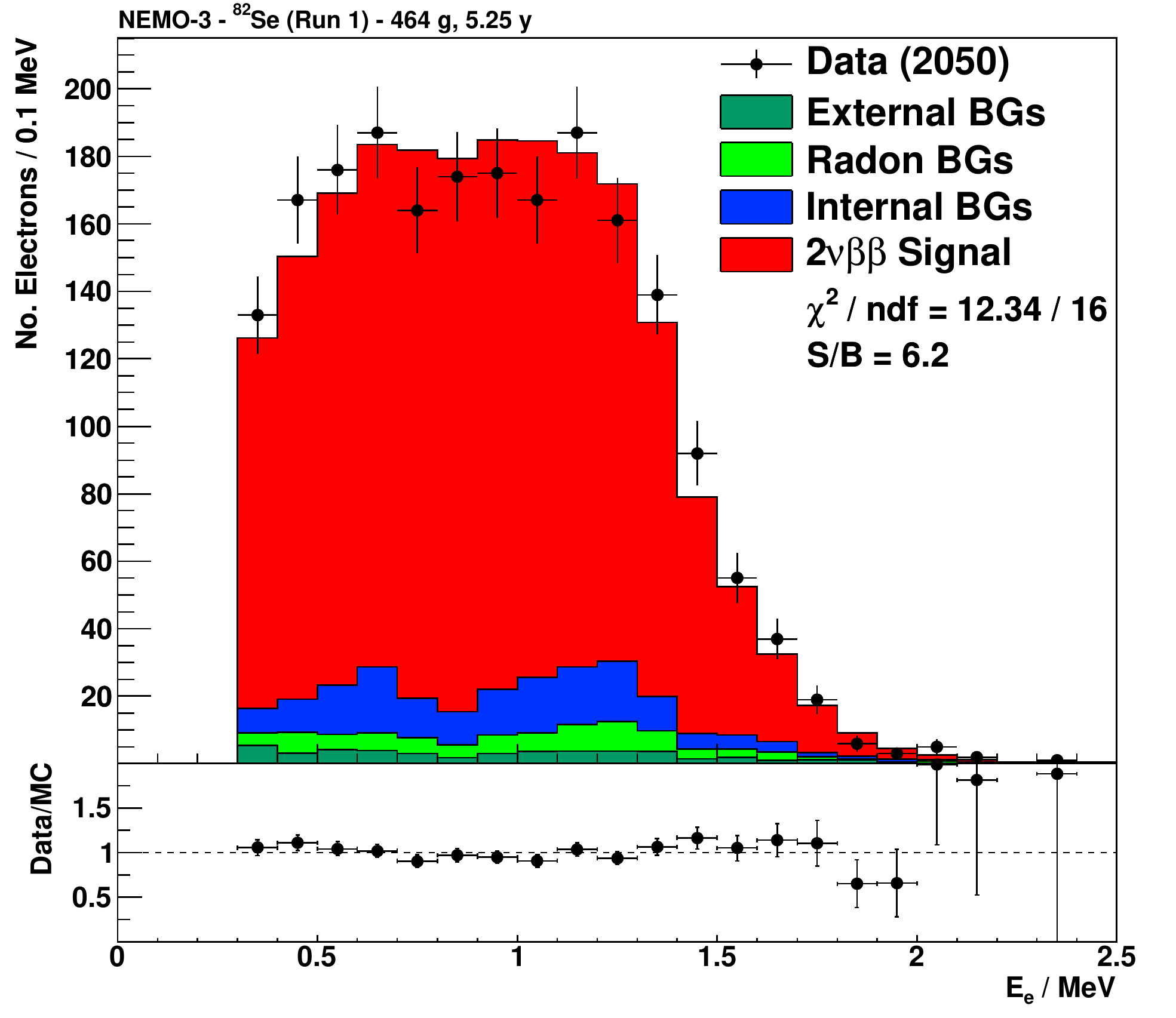} \label{fig:ESumCutSSD}}\\
  \caption{Energy distribution of individual electrons in the $\beta\beta$ channel for foils from enrichment run 1, after removing events where $\Sigma E_{e} < 1.6\,\mbox{MeV}$ to reduce the effect of contamination by $^{234\text{m}}$Pa.  The data are compared to the predicted spectrum from MC under the HSD and SSD hypotheses. There is good agreement between the data and SSD hypothesis ($\chi^{2}/\text{ndf} = 12.3/16$), but the HSD hypothesis is disfavoured ($\chi^{2}/\text{ndf} = 35.3/16$).}
  \label{fig:ESumCut}
\end{figure*}

The largest difference between the SSD and HSD single-electron energy spectra is at the low end of the 
distribution \cite{Simkovic:2005xx}. However, due to the previously identified issues with the 
$^{234\text{m}}$Pa conversion electron branching ratios, the individual electron energy distributions 
for the HSD and SSD models were compared to data after applying a cut on the sum of the electron energy 
of $E_{tot} > 1.6\,\mbox{MeV}$. This reduces the contamination from $^{234\text{m}}$Pa to below 2\%.
Figure~\ref{fig:ESumCut} shows a good agreement with data for the SSD hypothesis ($\chi^{2}/\text{ndf} = 12.3/16$)
while the HSD hypothesis is disfavoured ($\chi^{2}/\text{ndf} = 35.3/16$) at a level equivalent to 2.1$\sigma$. 
The SSD scenario is therefore assumed for the remainder of the analysis, unless explicitly stated otherwise.

\subsection{Extraction of $2\nu\beta\beta$ half-life}
\label{ssec:HalfLife}
\begin{figure}
  \includegraphics[width=0.5\textwidth]{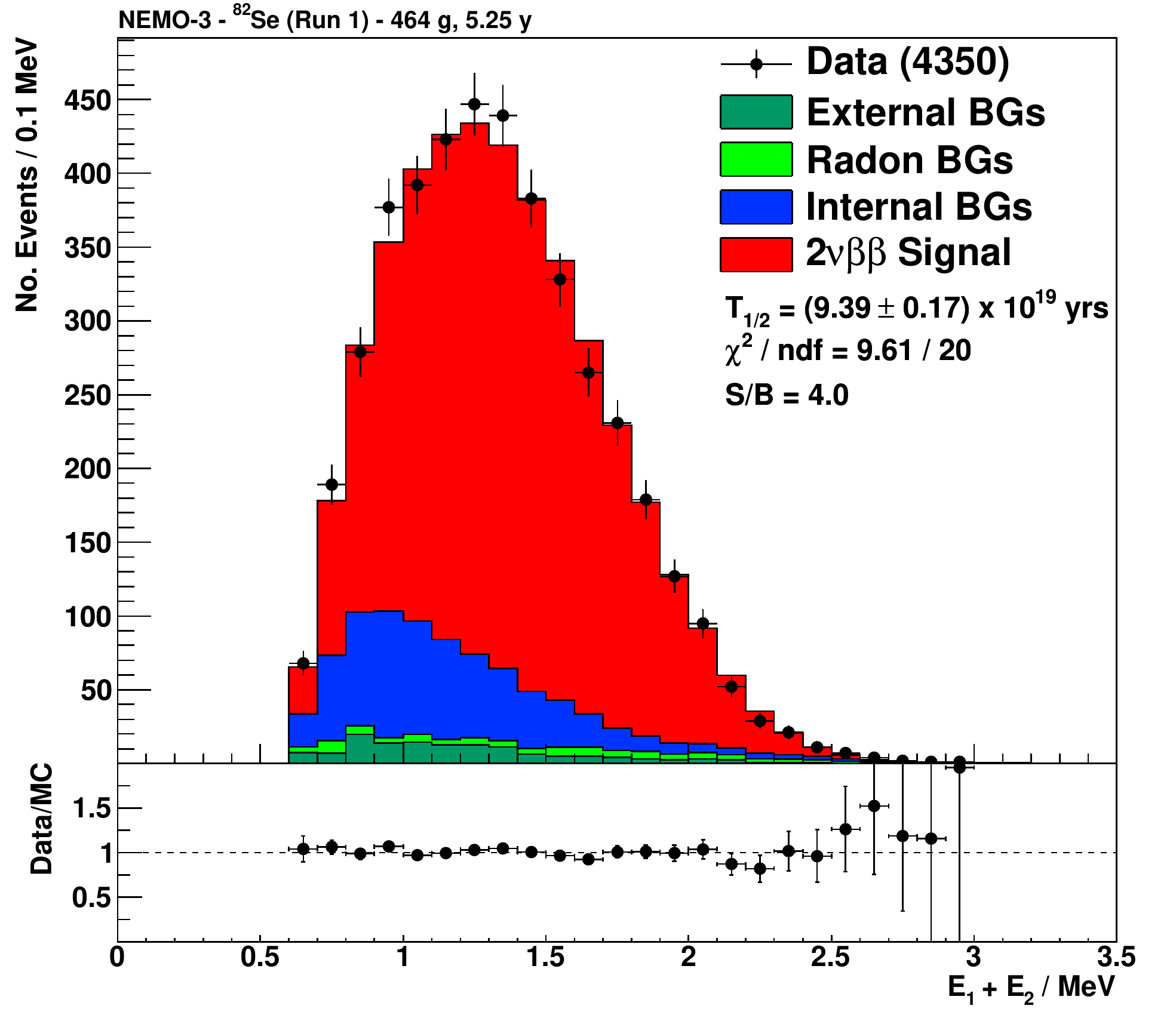} 
  \caption{Distribution of the summed energy of the two electrons in the $\beta\beta$ channel originating from enrichment run 1 foils. Data are compared to the MC prediction, where the activities of both signal and background components are taken from the binned log-likelihood fit.  The numbers of events in the histogram are as shown in Table~\ref{tab:TwoEEvents}.  The largest background category is internal contamination of the source foil (blue), but this is still much smaller than the contribution from the $2\nu\beta\beta$ signal, with a signal-to-background ratio of 4.0.}
\label{fig:twoEE}
\end{figure}

\begin{figure}
  \includegraphics[width=0.5\textwidth]{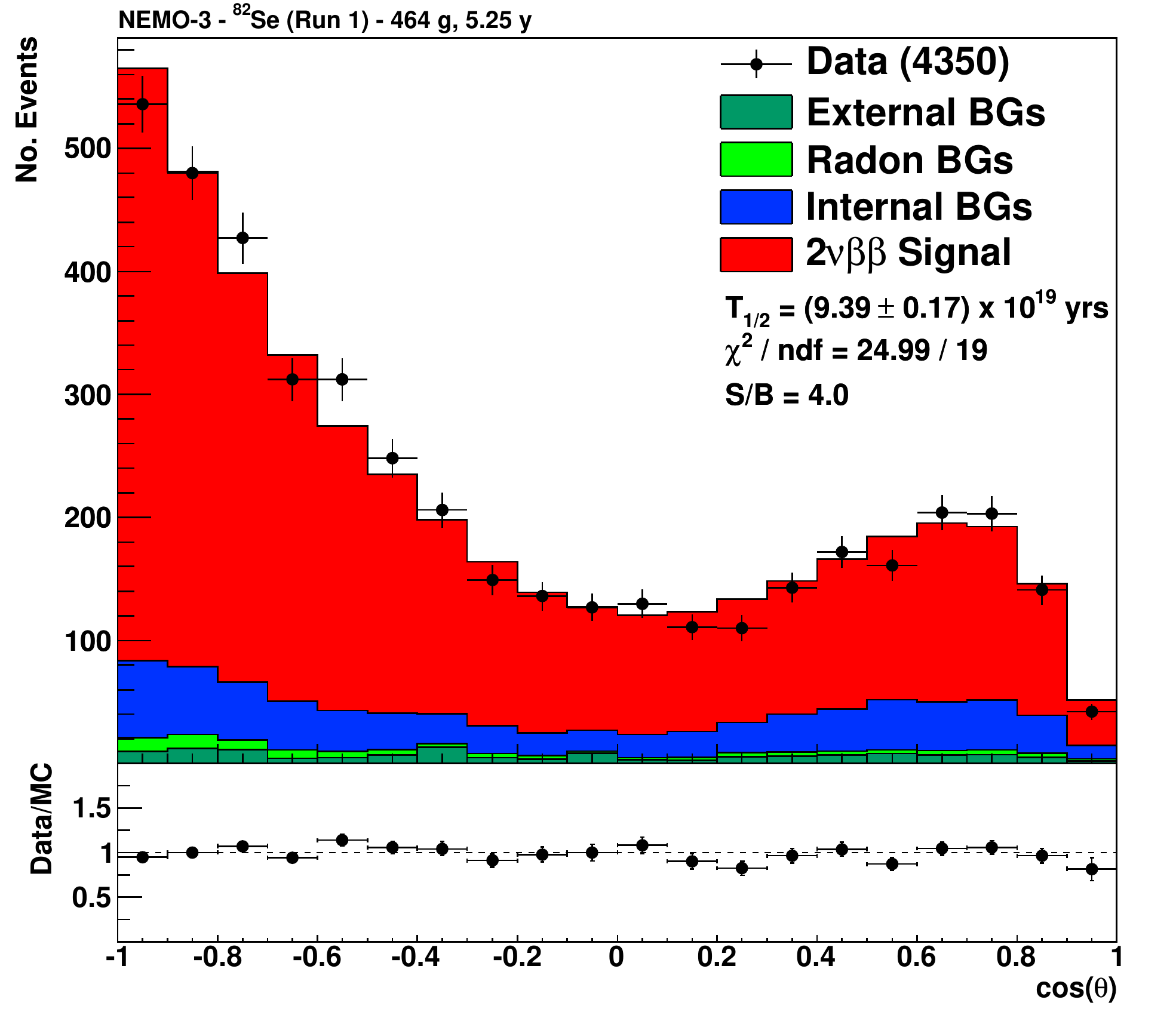}
  \caption{Distribution of the cosine of the angle between two electron tracks at the point of emission from the run 1 source foil in the $\beta\beta$ channel. As expected, more events are observed with electrons emitted to back-to-back than with smaller opening angles. This angular distribution has been reweighted based on data from $^{207}$Bi calibration sources.}
  \label{fig:TwoETwoNuAngle}
\end{figure}

A binned log-likelihood fit to the distribution of the sum of the two electron energies of the 4350 $\beta\beta$ events
selected from the data and originating from enrichment run 1 foils is performed together with a fit to the six background channels, as described in Section~\ref{sec:Backgrounds}.
The fit assuming the SSD hypothesis, shown in Figure~\ref{fig:twoEE}, yields $3472.4 \pm 75.7$ signal events, with a signal-to-background ratio of $4.0$.
The distribution of the opening angle between the two tracks is shown in Figure \ref{fig:TwoETwoNuAngle}.

In addition to the statistical uncertainty obtained from the log-likelihood fit, the $2\nu\beta\beta$ half-life measurement 
is affected by a number of systematic uncertainties. The most important source of systematic error 
is the uncertainty on the detector acceptance and reconstruction and selection efficiency.
This uncertainty is quantified using dedicated runs with $^{207}$Bi sources introduced into the detector 
and is compared with activities independently measured by an HPGe detector. Taking into account the 
systematic error on the HPGe measurement (5\%) the uncertainty on the signal efficiency 
is determined to be 5\% \cite{Arnold:2015wpy}.

Other sources of systematic uncertainty are listed in Table~\ref{tab:Systematics}. The systematic error 
due to the background modelling is dominated by the uncertainty on the $^{234\text{m}}$Pa conversion electron 
branching ratio discussed above. This uncertainty translates into a 2.3\% error on the $2\nu\beta\beta$ half-life 
for the run 1 foils and increases to 4.5\% if the analysis is performed on both enrichment samples
due to the higher $^{234\text{m}}$Pa levels in the run 2 foils. The uncertainty on the $2\nu\beta\beta$ half-life measurement 
is systematics dominated and therefore the overall precision of the measurement 
is improved by excluding the run 2 foils. 

The individual systematic errors are assumed to be uncorrelated and are added in quadrature 
to obtain the total systematic uncertainty of 6.3\%. This yields the final measurement of 
$N = 3472 \pm 76\,(\mbox{stat}) \pm 218\,(\mbox{syst})$
for the number of signal events obtained with $\left(0.464\pm0.002\right)\,\mbox{kg}$ of $^{82}$Se from enrichment run 1 over 5.25\,y
of observation. 

\begin{table}
  \caption{Systematic errors contributing to the uncertainty on the 2$\nu\beta\beta$ half-life measurement. The uncertainty on each source is given and its effect on the uncertainty on the 2$\nu\beta\beta$ half-life is shown.  The uncertainty for $^{234\text{m}}$Pa is shown for enrichment run 1 foils only and for enrichment runs 1 and 2 combined.}
  \label{tab:Systematics}
  \begin{tabular*}{\columnwidth}{@{\extracolsep{\fill}}lccc@{}}
    \hline\noalign{\smallskip}
                     & Systematic  & 2$\nu\beta\beta$ half-life\\
    Systematic cause & uncertainty & uncertainty\\
    \noalign{\smallskip}\hline\noalign{\smallskip}
    2$\nu\beta\beta$ efficiency   & $\pm 5.0\%$             & $\pm 5.0\%$  \\
    \multirow{2}{*}{$^{234\text{m}}$Pa modelling} & \multirow{2}{*}{$\pm 30.0\%$} & $\pm 2.3\%$ (Run 1)\\
                                  &                                                & $\pm 4.5\%$ (Runs 1+2)\\
    Min. $e^-$ energy             & $\left(0.3 - 0.8\right) \,\mbox{MeV}$ & $\pm 2.5\%$  \\
    Energy calibration            & $\pm 1.0\%$             & $\pm 1.25\%$ \\
    Int. BG activities            & $\pm 4.0\%$             & $\pm 0.8\%$  \\
    Ext. BG activities            & $\pm 10.0\%$            & $\pm 0.6\%$  \\
    Radon BG activities           & $\pm 10.0\%$            & $\pm 0.25\%$ \\
    $^{82}$Se mass                & $\pm 0.5\%$             & $\pm 0.5\%$  \\
    \noalign{\smallskip}\hline\noalign{\smallskip}
    \multirow{2}{*}{Total syst. error} & \multirow{2}{*}{N/A} & $\pm6.3\%$ (Run 1)\\
                                       &                      & $\pm7.3\%$ (Runs 1+2)\\ 
    \noalign{\smallskip}\hline
  \end{tabular*}
\end{table}

\noindent This can be converted to the $^{82}$Se $2\nu\beta\beta$ half-life using 
\begin{equation}
  T_{1/2} = \frac{\epsilon}{N}\frac{N_A m}{A} \ln{(2)} t \,,
\end{equation}
where $\epsilon$ is the selection efficiency (2.623\%), $N_A$ is Avogadro's number, 
$\frac{m}{A}$ is the number of moles of $^{82}$Se and $t$ is the total exposure time. 
The resulting half-life, assuming the SSD hypothesis, is
\begin{equation}
  T_{1/2}^{2\nu} = \left[9.39 \pm 0.17\,(\mbox{stat}) \pm 0.58\,(\mbox{syst})\right] \times 10^{19}\,\mbox{y} \,.
\end{equation}
An identical analysis under the HSD hypothesis gives
\begin{equation}
  T_{1/2}^{2\nu} = \left[10.63 \pm 0.19\,(\mbox{stat}) \pm 0.66\,(\mbox{syst})\right] \times 10^{19}\,\mbox{y} \,.
\end{equation}

The half-life measurement allows the experimental determination of the NME for the $2\nu\beta\beta$  
decay mode of $^{82}$Se using the equation
\begin{equation}
  \left(T_{1/2}^{2\nu}\right)^{-1} = G^{2\nu}\left(Q_{\beta\beta},Z\right) g^{4}_{A} \left|M^{2\nu}\right|^2 \,,
\end{equation}
where $g_A $ is the axial-vector coupling constant and $G^{2\nu}$ is the phase
space for the $^{82}$Se $2\nu\beta\beta$ $0^+ \rightarrow 0^+$ ground state transition.
Taking $G^{2\nu}\left(Q_{\beta\beta},Z\right) = 1.6 \times 10^{-18}\,\mbox{y}^{-1}$ as calculated in
\cite{Kotila:2012zza,Mirea:2015nsl} and assuming $g_A = 1.27$ \cite{Agashe:2014kda} we obtain for the matrix element under the SSD hypothesis
\begin{equation}
  \left|M^{2\nu}\right| = 0.0498 \pm 0.0016 \,,
\end{equation}
and under the HSD hypothesis
\begin{equation}
  \left|M^{2\nu}\right| = 0.0468 \pm 0.0015 \,,
\end{equation}
where the quoted errors include both statistical and systematic uncertainties, which are assumed to be uncorrelated.

\section{Neutrinoless double beta decay}
\label{sec:TwoEZeroNu}

A search for $0\nu\beta\beta$ is carried out by selecting $\beta\beta$ events as outlined in Section~\ref{sec:ParticleID}.
Due to the higher energies of electrons emitted in the $0\nu\beta\beta$ decay the uncertainties due to 
the $^{234\text{m}}$Pa background model reported earlier are negligible. 
Consequently, both enrichment samples are included in the $0\nu\beta\beta$ analysis. 
Alongside backgrounds from natural radioactivity,
$0\nu\beta\beta$ has an additional background contribution from $2\nu\beta\beta$ events.  The following results assume the SSD
hypothesis, but the same results are also found if the HSD case is taken.
We considered four lepton number violating mechanisms for $0\nu\beta\beta$: light Majorana neutrino exchange, 
the admixture of right-handed currents in electroweak interactions, $0\nu\beta\beta$ decay accompanied by a majoron emission
and R-parity violating SUSY models.  No evidence for a $0\nu\beta\beta$ signal is found for any of these mechanisms and therefore 
corresponding limits on the half-lives are set. 
The background contributions to $0\nu\beta\beta$ in the [$2.6-3.2$]\,MeV energy region, where
most of the signal from  the light Majorana neutrino exchange and right-handed current mechanisms is expected, 
are shown in Table~\ref{tab:TwoEEvents}. 

The electron energy sum distribution is used to set the limits
using a modified frequentist method based on a binned log-likelihood ratio test statistic (CL$_{s}$) \cite{Junk:1999kv}.  
The statistic is calculated over the entire energy range above 0.6\,MeV 
and takes into account the shape of the energy distribution. 

In order to estimate the effect of systematic uncertainties on the limit,  the background and signal distributions are scaled 
by random factors drawn from Gaussian distributions with widths defined by the systematic errors of the experiment \cite{Fisher:2006zz}, which are given 
in Table~\ref{tab:ZeroNuSystematics}. Similarly to $2\nu\beta\beta$, the most significant contribution comes from 
the error on the selection efficiency.

\begin{table}
  \caption{Values of the $1\sigma$ systematic errors included when setting limits on 0$\nu\beta\beta$ decay modes using the \texttt{COLLIE} software package \cite{Fisher:2006zz}. The estimated errors shown are on the systematic quantity and are therefore significantly reduced when transferred through the $\beta\beta$ selection, with the exception of the dominating $0\nu\beta\beta$ efficiency uncertainty.}
  \label{tab:ZeroNuSystematics}
  \begin{tabular*}{\columnwidth}{@{\extracolsep{\fill}}lcc@{}}
    \hline\noalign{\smallskip}
                     & Systematic  & \\
    Systematic cause & uncertainty & Source of estimate\\
    \noalign{\smallskip}\hline\noalign{\smallskip}
    0$\nu\beta\beta$ efficiency         & $\pm5.0\%$                   & $^{207}$Bi vs. HPGe\\
    \multirow{2}{*}{Ext. BG activities} & \multirow{2}{*}{$\pm10.0\%$} & Variation from\\
                                        &                              & model in \cite{Argyriades:2009vq}\\
    Radon BG activities                 & $\pm10.0\%$                  & 1e1$\alpha N \gamma$ vs. 1e1$\gamma$ \\
    Int. BG activities                  & \multirow{2}{*}{$\pm4.0\%$}  & \multirow{2}{*}{$^{207}$Bi $1eN\gamma$ vs. 2e} \\
    (excl. Tl, Bi \& Pa)\\
    Int. $^{214}$Bi activity            & $\pm10.0\%$                  & 1e1$\alpha N \gamma$ vs. 1e1$\gamma$ \\
    Int. $^{208}$Tl activity            & $\pm10.0\%$                  & $^{232}$U vs. HPGe \\
    Int. $^{234\text{m}}$Pa activity    & $\pm30.0\%$                  & Old vs. new MC\\
    2$\nu\beta\beta$ activity           & $\pm1.0\%$                   & Statistical error\\
    \noalign{\smallskip}\hline
  \end{tabular*}
\end{table}

\subsection{Light Majorana Neutrino Exchange}
\label{sec::MMResults}

Light Majorana neutrino exchange is the most commonly discussed mechanism of $0\nu\beta\beta$ decay.
It has an experimental signature characterised by a peak in the distribution of the electron energy sum at the $Q_{\beta\beta}$ value. 

\begin{figure*}
  \subfloat[Light Neutrino Exchange]                       {\includegraphics[width=0.5\textwidth]{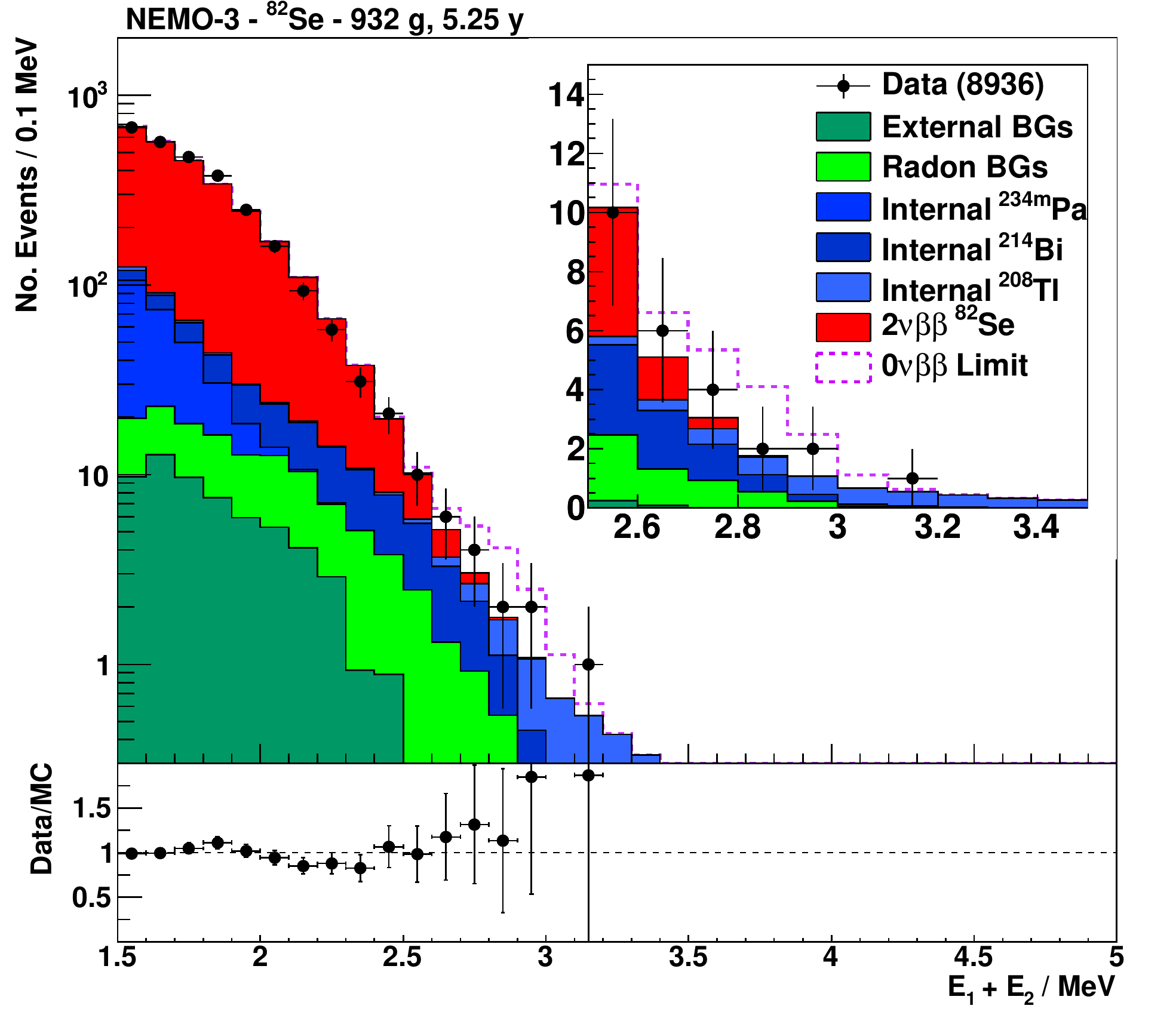}  \label{fig:MMLimit}}\hfill%
  \subfloat[Right-handed Current $\langle \lambda \rangle$]{\includegraphics[width=0.5\textwidth]{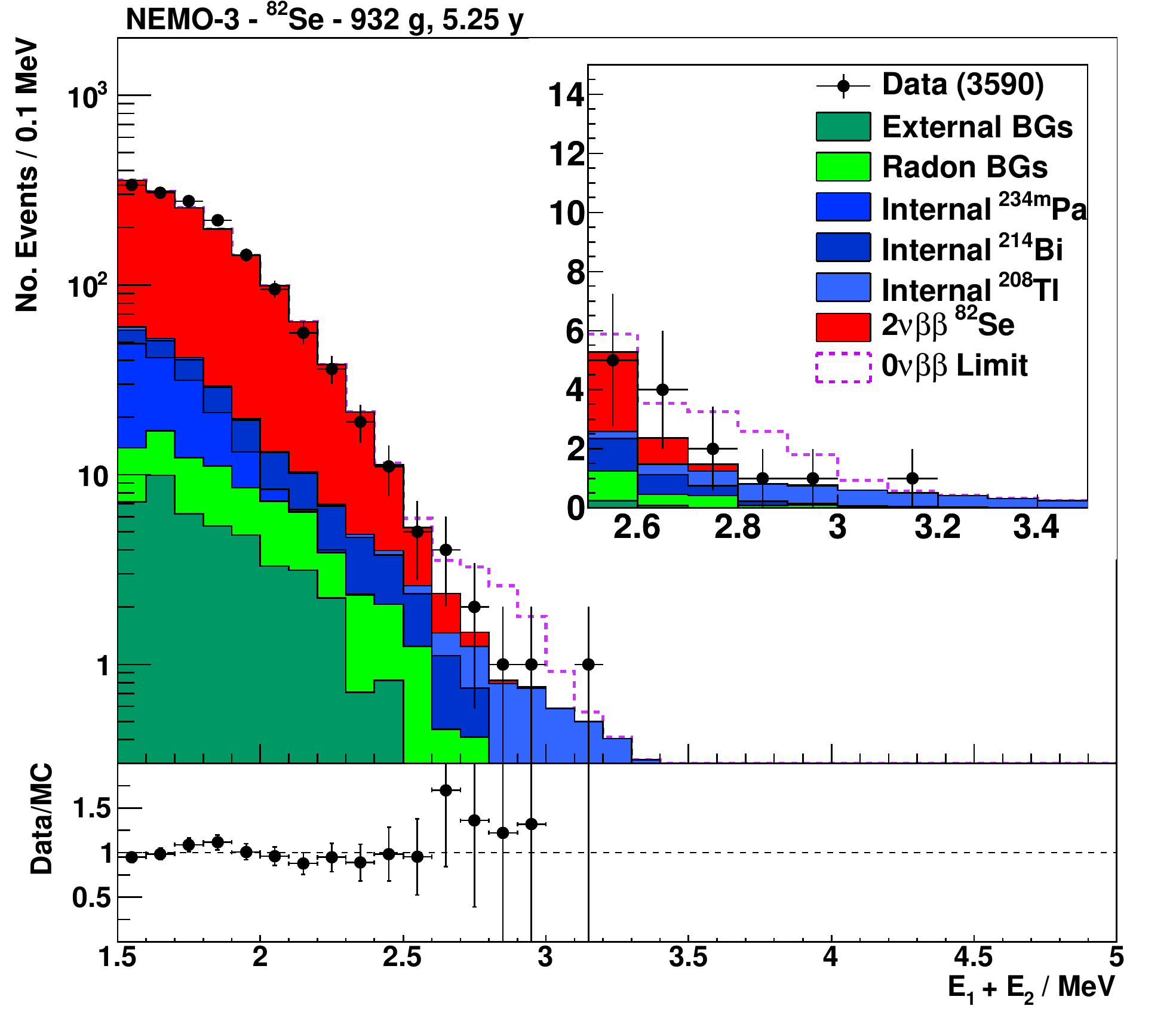}\label{fig:RHCLimit}}\\
  \caption{Distribution of the summed electron energies in the $\beta\beta$ channel and the ratio between the observed and MC predicted data.  The inset plot shows the highest energy events on a linear scale.  The solid histograms represent the backgrounds and $2\nu\beta\beta$ predictions and the open histogram shows a hypothetical 0$\nu\beta\beta$ signal corresponding to the limit at 90\% C.L. Figure~(a) contains events selected in the $\beta\beta$ channel and Figure~(b) contains a subset of these events that also pass the energy asymmetry cut for the right-handed current $\langle \lambda \rangle$ mode, $A>0.26$, where $A$ is defined in the text.}
\end{figure*}

The  background, signal and data distributions shown in Figure~\ref{fig:MMLimit} are used to set the limit.
There are 7.20 [$5.09-10.66$] events expected to be excluded at the 90\% C.L., where the $\pm1\sigma$ range is given in brackets.
The systematic errors from Table~\ref{tab:ZeroNuSystematics} are included in the expected limit and only reduce it by 2\%.
Taking into account the detector efficiency of 9.80\% for this $0\nu\beta\beta$ mechanism and the $^{82}$Se exposure of 
4.90\,kg$\cdot$y, the 90\% C.L. expected half-life limit is $3.39\, [2.29 - 4.80] \times 10^{23}\,\mbox{y}$.
From the data sample, 9.67 events are excluded at 90\% C.L. leading to an upper limit on the half-life of 
\begin{equation}
  T_{1/2}^{0\nu} > 2.5 \times 10^{23} \,\mbox{y} \,(90\%\,\mbox{C.L.}) \,,
\end{equation}
which is within the $1 \sigma$ range of the expected sensitivity. 

Equation~\ref{equ:0vbb} is used to convert the half-life limit into an upper bound on the effective Majorana neutrino mass. 
The phase space is taken as $G^{0\nu} = 1.016 \times 10^{-14} \,\mbox{y}^{-1}$ \cite{Kotila:2012zza} (in agreement with
$G^{0\nu} = 1.014 \times 10^{-14} \,\mbox{y}^{-1}$ from \cite{Mirea:2015nsl}).

Several nuclear models are used to calculate the NME for the $^{82}$Se $0\nu\beta\beta$ transition to the ground state. 
The most recent calculations from \cite{Menendez:2009jp,Simkovic:2013qiy,Suhonen:2015rh,Iachello:2015hwa,Rodriguez:2010mn,Yao:2015new} 
have been used and $g_A$ is taken in the range $1.25-1.27$ to correspond with the assumptions of the different calculations.
As a result, the constraint on the effective neutrino mass is
\begin{equation}
  \langle m_{\nu} \rangle < \left(1.2 - 3.0\right) \,\mbox{eV} \,(90\% \,\mbox{C.L.}) \,.
\end{equation}

\subsection{Right-handed Currents}
\label{sec::RHCResults}

Right-left symmetric models can provide an alternative mechanism for $0\nu\beta\beta$ due to the presence of right-handed 
currents (RHC) in the electroweak Lagrangian \cite{Doi:1985dx,Tomoda:1990rs}. The lepton number violation mechanism is characterised by 
the coupling between right-handed currents of quarks and leptons, $\langle \lambda \rangle$,
and right-handed quark and left-handed lepton currents, $\langle \eta \rangle$. 

The $\langle \lambda \rangle$ mechanism leads to very different angular and single energy distributions 
of the final state electrons and can therefore be distinguished from other mechanisms in an experiment capable of reconstructing 
the full topology of the process, such as NEMO-3 \cite{Arnold:2010tu}.  In addition to the electron energy sum, further 
discrimination between the RHC $\langle \lambda \rangle$ mechanism and background can be achieved with the energy asymmetry
between the individual electron energies, $A$, defined as
$A = \left( E_{\text{max}} - E_{\text{min}} \right) / \left( E_{\text{max}} + E_{\text{min}} \right)$.

The expected sensitivity in the RHC $\langle \lambda \rangle$ mode has been studied by MC and is maximised with a cut
of $A>0.26$. This selection is therefore applied when searching for this particular decay mode as shown in Figure~\ref{fig:RHCLimit}. 
Cutting on the energy asymmetry variable provides no improvement in sensitivity for the $\langle \eta \rangle$ mode
and so the standard $\beta\beta$ selection criteria are used in this case. 

For the $\langle \lambda \rangle$ mode, 7.34 events are excluded from the data sample leading to a lower limit
on the half-life of $1.63 \times 10^{23}$\,y at 90\% C.L. This result is in agreement with the median expected
sensitivity of the experiment of $2.16~[1.46 - 3.01] \times 10^{23}$\,y. For the $\langle \eta \rangle$ mode,
the half-life lower limit is $2.19 \times 10^{23}$\,y at 90\% C.L. and also agrees with the expected sensitivity. 

These half-life limits are translated into upper bound on the coupling between right-handed quark and lepton currents, 
$\langle \lambda \rangle < (2.2 - 2.6) \times 10^{-6}$, and into the coupling between right-handed quark 
and left-handed lepton currents, $\langle \eta \rangle < (1.7 - 2.1) \times 10^{-8}$.
The constraints are obtained using NME calculations from \cite{Aunola:1998jc,Muto:1989cd,Tomoda:1990rs}.

\subsection{Majoron Emission}
\label{sec::MajoronResults}

A $0\nu\beta\beta$ decay accompanied by a majoron, a light or massless boson that weakly couples to the neutrino, 
has a continuous spectrum of the energy sum of the two decay electrons, $E_{tot}$, up to $Q_{\beta\beta}$ \cite{Bamert:1994hb}. 
The phase space of the process depends on the spectral index $n$, as 
$G^{0\nu} \propto \left( Q_{\beta\beta} - E_{tot} \right)^n$, and determines the shape of the distribution. 
Decays with higher $n$ have broader $E_{tot}$ distributions peaking at lower energy values. 
Such events are harder to separate from $2\nu\beta\beta$ and other backgrounds. Therefore only the result of the search 
for majoron induced $0\nu\beta\beta$ decay with $n=1$ is shown here.
The corresponding half-life limit is $T^{0\nu}_{1/2} > 3.7 \times 10^{22}$\,y at 90\% C.L., which translates 
into an upper limit on the majoron-neutrino coupling of $\langle g_{ee} \rangle < (3.2-8.0) \times 10^{-5}$.
The range is due to a spread in NME calculations, which are taken from 
\cite{Menendez:2009jp,Simkovic:2013qiy,Suhonen:2015rh,Iachello:2015hwa,Rodriguez:2010mn,Yao:2015new},
while the phase space is taken from \cite{Kotila:2015ata}.

\subsection{Supersymmetry Models}
\label{sec:SUSYResults}
R-parity violating SUSY models can trigger 0$\nu\beta\beta$ decay via short range exchange of 
heavy superpartners, such as gluino or neutralino, or long range exchange of squarks and neutrinos \cite{Faessler:1997db,Faessler:2007nz}.  
The kinematics of the electrons emitted in the decay are the same as in the light neutrino exchange mechanism
and therefore the same half-life limit can be used to set limits on SUSY parameters. 
Taking the phase space from \cite{Kotila:2012zza} and the NME from \cite{Faessler:2011xx,Faessler:2011yy}, the following constraints 
are obtained for the short range gluino and neutralino exchange mechanisms:
\begin{equation}
\lambda^{\prime}_{111} \leq (7.68 - 8.32) \times 10^{-2} \left(\frac{m_{\tilde q}}{1~\text{TeV}}\right)^2 \left(\frac{m_{\tilde g}}{1~\text{TeV}}\right)^{1/2} \,,
\end{equation}
\begin{equation}
\lambda^{\prime}_{111} \leq (5.33 - 5.78) \times 10^{-1} \left(\frac{m_{\tilde e}}{1~\text{TeV}}\right)^2 \left(\frac{m_{\tilde \chi}}{1~\text{TeV}}\right)^{1/2},
\end{equation}
where $m_{\tilde q}$, $m_{\tilde g}$, $m_{\tilde e}$ and  $m_{\tilde \chi}$ are the masses of squark, gluino, selectron and neutralino respectively. 
The corresponding limits for the long range squark exchange mechanism are:
\begin{equation}
 \lambda^{\prime}_{111} \lambda^{\prime}_{111} \leq (3.17 - 3.22) \times 10^{-2} \left(\frac{\Lambda_{\rm SUSY}}{1~\text{TeV}}\right)^3 \,,
 \end{equation}
  \begin{equation}
  \lambda^{\prime}_{112} \lambda^{\prime}_{121} \leq (1.66 - 1.68) \times 10^{-3} \left(\frac{\Lambda_{\rm SUSY}}{1~\text{TeV}}\right)^3 \,,
 \end{equation} 
\begin{equation} 
  \lambda^{\prime}_{113} \lambda^{\prime}_{131} \leq  (6.88 - 6.98)  \times 10^{-5} \left(\frac{\Lambda_{\rm SUSY}}{1~\text{TeV}}\right)^3 \,,
\end{equation} 
where $\Lambda_{\rm SUSY}$ is a general SUSY breaking scale parameter. The above limits assume $g_A = 1.25$. The spread in the limits is
due to NME uncertainties associated with differences in the form of the Argonne and Charge Dependent Bonn (CD-Bonn) nucleon-nucleon
potentials \cite{Faessler:2011xx}. 

\section{Summary and Conclusions}
The results of $^{82}$Se $\beta\beta$ decay studies obtained with the full set of NEMO-3 data are presented. 
The $^{82}$Se $2\nu\beta\beta$ decay half-life for the ground state transition has been measured using foils from
the first enrichment run only, due to higher levels of $^{234\text{m}}$Pa contamination in the foils from the second run
and associated uncertainties in the $^{234\text{m}}$Pa conversion electron branching ratios. 
With the corresponding exposure of 2.4\,kg$\cdot$y, the HSD transition hypothesis is disfavoured at the 2.1$\sigma$ level,
whilst the SSD hypothesis is supported.  In the SSD scenario, the half-life has been measured to be
$T_{1/2}^{2\nu} = \left[9.39 \pm 0.17\,(\mbox{stat}) \pm 0.58\,(\mbox{syst})\right] \times 10^{19}\,\mbox{y}$.  This is the most precise
measurement for this isotope to date and allows the experimental extraction of the corresponding
NME, $ \left|M^{2\nu}\right| = 0.0498 \pm 0.0016$.  This single result is more precise than and consistent
with the world average reported in \cite{Saakyan:2013rvw, Barabash:2015eza}.
The SuperNEMO experiment is based on the same design principles as the NEMO-3 detector and will have lower backgrounds and improved energy resolution.
A demonstrator module is currently being commissioned, which will house 7\,kg of $^{82}$Se.  
The SuperNEMO demonstrator module  will have the sensitivity to
distinguish between the SSD and HSD scenarios at a $>5\sigma$ level. 

A search for $0\nu\beta\beta$ decay has been carried out for a number of different mechanisms, with foils from both enrichment runs, giving an exposure
of 4.9\,kg$\cdot$y. No evidence for any neutrinoless double beta decay transition is found and therefore upper limits on the corresponding
lepton number violating parameters have been set. The results of the $0\nu\beta\beta$ search are summarised in Table~\ref{tab:Limits}.
The most stringent half-life limit for $^{82}$Se is obtained for the light neutrino exchange mechanism of $0\nu\beta\beta$,
$T_{1/2}^{0\nu} > 2.5 \times 10^{23} \,\mbox{y}$ at 90\% C.L. corresponding to an effective Majorana neutrino mass of
$\langle  m_{\nu} \rangle < \left(1.2 - 3.0\right) \,\mbox{eV}$.  It should be noted that the CUPID-0 collaboration recently published their first limit for $0\nu\beta\beta$  of $^{82}$Se with a value $T_{1/2}^{0\nu} > 2.4 \times 10^{24}\,\mbox{y}$ \cite{Azzolini:2018dyb}.

\begin{table*}
  \caption{Limits from 0$\nu\beta\beta$ searches for different decay modes in $^{82}$Se. The signal efficiency and the 90\% C.L. limits for half-lives and lepton number violating (LNV) parameters are shown. The ranges in the expected half-life limits are the $\pm1\sigma$ range of systematic uncertainties on the background model and signal efficiency.  The ranges in the LNV parameter are due to the spread in NME calculations. The R-parity violating SUSY LNV parameters correspond to sparticle masses and energy scale $\Lambda_{\rm SUSY}$ in TeV.}
  \label{tab:Limits}
  \begin{tabular*}{\textwidth}{@{\extracolsep{\fill}}lccccc@{}}
    \hline\noalign{\smallskip}
    \multirow{2}{*}{0$\nu\beta\beta$ mechanism} & \multirow{2}{*}{Mode} & Efficiency & \multicolumn{2}{c}{$T_{1/2}^{0\nu}$ 90\% C.L. $\left(10^{23}\,\mbox{y}\right)$} & LNV parameter\\
                                                &                       & (\%) & Expected       & Observed                                                    & 90\% C.L. \\
    \noalign{\smallskip}\hline\noalign{\smallskip}
    Light neutrino exchange                     & $\langle m_{\nu} \rangle$         & 9.80  & $>2.29 - 4.80$ & $>2.53$ & $<\left(1.2-3.0\right)$\,eV \\ 
    \noalign{\smallskip}\hline\noalign{\smallskip}
    \multirow{2}{*}{Right-handed currents}      & $\langle \lambda \rangle$         & 4.79 & $>1.46 - 3.01$ & $>1.63$ & $<(2.2-2.6) \times 10^{-6}$ \\ 
                                                & $\langle \eta \rangle $           & 8.70 & $>1.99 - 4.20$ & $>2.19$ & $<(1.7-2.1) \times 10^{-8}$ \\ 
    \noalign{\smallskip}\hline\noalign{\smallskip}
    Majoron emission                            & $\langle g_{ee} \rangle _{n=1}$   & 7.03 & $>0.23 - 0.48$ & $>0.37$ &  $<(3.2-8.0) \times 10^{-5} $  \\ 
    \noalign{\smallskip}\hline\noalign{\smallskip}
    \multirow{5}{*}{R-parity violating SUSY}    & $\lambda^{\prime}_{111}$          & \multirow{5}{*}{9.80} & \multirow{5}{*}{$>2.29 - 4.80$} & \multirow{5}{*}{$>2.53$} & $\leq (7.68 - 8.32) \times 10^{-2} m_{\tilde q}^2  m_{\tilde g}^{1/2}$\\
                                                & $\lambda^{\prime}_{111}$          &                      &                &          & $\leq (5.33 - 5.78) \times 10^{-1} m_{\tilde e}^2 m_{\tilde \chi}^{1/2}$\\
                                                & $\lambda^{\prime}_{111} \lambda^{\prime}_{111}$ &                      &                &          & $\leq (3.17 - 3.22) \times 10^{-2} \Lambda_{\rm SUSY}^3$\\
                                                & $\lambda^{\prime}_{112} \lambda^{\prime}_{121}$ &                      &                &          & $\leq (1.66 - 1.68) \times 10^{-3} \Lambda_{\rm SUSY}^3$\\
                                                & $\lambda^{\prime}_{113} \lambda^{\prime}_{131}$ &                      &                &          & $\leq  (6.88 - 6.98)  \times 10^{-5} \Lambda_{\rm SUSY}^3$\\
    \noalign{\smallskip}\hline
  \end{tabular*}
\end{table*}

The constraints on the RHC parameters, $\langle \lambda \rangle$ and $\langle \eta \rangle$, on 
the majoron-neutrino coupling constant, $\langle g_{ee} \rangle$, and on R-parity violating SUSY parameters, 
$\lambda^{\prime}_{1ij}$, shown in Table~\ref{tab:Limits} are the best for $^{82}$Se and are comparable with the
best available limits from other isotopes \cite{Arnold:2015wpy} despite a much lower exposure. 

\begin{acknowledgement}
We thank the staff of the Modane Underground Laboratory for their technical assistance in running the experiment. We acknowledge support by the grants agencies of the Czech Republic, CNRS/IN2P3 in France, RFBR in Russia, STFC in the UK and NSF in the USA.
\end{acknowledgement}

\bibliography{NEMO3_Se82}

\end{document}